\documentclass[11pt,a4paper]{article}
\pdfoutput=1
\usepackage{jheppub}
\usepackage{tikz}
\usepackage{subcaption}
\DeclareMathOperator{\Tr}{Tr}

\newcommand{\ri}{\mathrm{i}}

\newcommand{\Th}{\Theta}
\newcommand{\cob}{\delta}
\newcommand{\ep}{\epsilon}

\newcommand{\hf}{\frac{1}{2}}
\newcommand{\qu}{\frac{1}{4}}
\newcommand{\til}[1]{\widetilde{#1}}
\newcommand{\si}{\sigma}

\renewcommand{\b}[1]{\overline{#1}}
\newcommand{\del}{\partial}

\newcommand{\bra}{\langle}
\newcommand{\ket}{\rangle}
\newcommand{\la}{\lambda}

\newcommand{\ka}{\kappa}
\newcommand{\h}[1]{\widehat{#1}}
\newcommand{\bt}{\beta}
\newcommand{\ga}{\gamma}
\newcommand{\Ga}{\Gamma}
\newcommand{\al}{\alpha}

\newcommand{\rt}[1]{\sqrt{#1}}
\newcommand{\cO}{\mathcal{O}}
\newcommand{\cZ}{\mathcal{Z}}
\newcommand{\cF}{\mathcal{F}}
\newcommand{\cR}{\mathcal{R}}

\newcommand{\cM}{\mathcal{M}}
\newcommand{\tB}{\tilde{B}}
\newcommand{\tI}{\tilde{I}}

\begin{document}

\title{JT supergravity and Brezin-Gross-Witten tau-function}

\author[a]{Kazumi Okuyama}
\author[b]{and Kazuhiro Sakai}

\affiliation[a]{Department of Physics, Shinshu University,\\
3-1-1 Asahi, Matsumoto 390-8621, Japan}
\affiliation[b]{Institute of Physics, Meiji Gakuin University,\\
1518 Kamikurata-cho, Totsuka-ku, Yokohama 244-8539, Japan}

\emailAdd{kazumi@azusa.shinshu-u.ac.jp, kzhrsakai@gmail.com}

\abstract{
We study thermal correlation functions of Jackiw-Teitelboim (JT)
supergravity. We focus on the case of JT supergravity on orientable
surfaces without time-reversal symmetry.
As shown by Stanford and Witten recently,
the path integral amounts to the computation of the volume of
the moduli space of super Riemann surfaces, which is characterized
by the Brezin-Gross-Witten (BGW) tau-function of the KdV hierarchy.
We find that the matrix model of JT supergravity
is a special case of the BGW model with infinite number of couplings 
turned on in a specific way, by analogy with the relation between
bosonic JT gravity and the Kontsevich-Witten (KW) model.
We compute the genus expansion of the one-point function of
JT supergravity and study its low-temperature behavior.
In particular, we propose a non-perturbative completion of the
one-point function in the Bessel case where all couplings
in the BGW model are set to zero.
We also investigate the free energy and correlators
when the Ramond-Ramond flux is large.
We find that by defining a suitable basis higher genus free energies
are written exactly in the same form as those of the KW model,
up to the constant terms coming from the volume of the unitary group.
This implies that the constitutive relation of the KW model is
universal to the tau-function of the KdV hierarchy.
}

\maketitle

%%%%%%%%%%%%%%%%%%%%%%%%%%%%%%%%%%%%%%%%%%%%%%%%%%%%%%%%%%%%%%%%%%%%%%%%

\section{Introduction}
Jackiw-Teitelboim (JT) gravity \cite{Jackiw:1984je,Teitelboim:1983ux} 
is a simple model of 2d gravity coupled to
a scalar field, which serves as a solvable example of the 
AdS${}_2$/CFT${}_1$ correspondence
\cite{Almheiri:2014cka,Maldacena:2016upp,Engelsoy:2016xyb,Harlow:2018tqv}.
In a recent paper \cite{Saad:2019lba}, Saad, Shenker, and Stanford showed that
the sum over topologies in JT gravity is reproduced from a matrix model,
based on the observation of Eynard and Orantin \cite{Eynard:2007fi} that the 
recursion relation for the Weil-Petersson volume found by 
Mirzakhani \cite{mirzakhani2007simple} can be recast as a topological recursion 
for a certain double-scaled matrix model \cite{Eynard:2007kz}.
This clearly shows that JT gravity on asymptotically AdS${}_2$ is holographically dual to
an ensemble of one-dimensional quantum systems, where the Hamiltonian of
a quantum system plays the role of a random matrix.

In our previous papers \cite{Okuyama:2019xbv,Okuyama:2020ncd}, by generalizing the
method of Zograf \cite{Zograf:2008wbe},
we have developed a systematic technique of computing the genus expansion
of the connected correlators $\bra Z(\bt_1)\cdots Z(\bt_n)\ket_c$ of the thermal partition functions $Z(\bt_i)=\Tr e^{-\bt_i H}$ in the JT gravity matrix model. In particular, we found that the JT gravity matrix model
is nothing but a special case of the Kontsevich-Witten (KW) topological gravity 
\cite{Kontsevich:1992ti,Witten:1990hr} where
infinite number of couplings are turned on with a specific value.

In this paper, we will show that a similar story holds for JT supergravity as well.
In a paper by Stanford and Witten \cite{Stanford:2019vob},
it was shown that the matrix model description in \cite{Saad:2019lba} can be generalized to
JT supergravity and classified by the Altland-Zirnbauer ensemble
depending on the symmetry of the system. In this paper we will focus on the case of 
JT supergravity on orientable surfaces without time-reversal symmetry. 
As discussed in \cite{Stanford:2019vob,norbury2,Witten}, in this case
the corresponding matrix model computes the volume of 
the moduli space of super Riemann surfaces and it is characterized by the 
Brezin-Gross-Witten (BGW) $\tau$-function of the KdV hierarchy.
We find that the matrix model of JT supergravity
is a special case of the BGW model with infinite number of couplings turned on
in a specific way, which is almost parallel to the case of bosonic JT gravity
by replacing the KW model by the BGW model.

We first compute the genus expansion of JT supergravity using the cut-and-join
representation of the BGW $\tau$-function \cite{Alexandrov:2016kjl} and
study the low temperature behavior of the one-point function $\bra Z(\bt)\ket$.
Next, we consider
the regime in which the Ramond-Ramond (RR) flux $\nu$ is large:
\begin{equation}
\begin{aligned}
% \nu\to\infty,~~\hbar\to0\quad\text{with}\quad q=\hbar\nu:~\text{fixed},
\nu\gg 1,~~\hbar\ll 1\quad\text{with}\quad q=\hbar\nu:~\text{fixed},
\end{aligned} 
\label{eq:thooft-limit}
\end{equation}
where $\hbar$ is the genus-counting parameter.
In this regime we show that the higher genus free energy of 
the BGW model is written in terms of the genus-zero quantities. This kind of relation is 
known as the constitutive relation \cite{Dijkgraaf:1990nc,Eguchi:1994cx}. This enables us to study the
genus expansion of connected correlators $\bra Z(\bt_1)\cdots Z(\bt_n)\ket_c$
by the standard technique of the old matrix model 
(see \cite{Ginsparg:1993is} for a review). 

This paper is organized as follows. In section \ref{sec:JT}, we show that the matrix model
of JT supergravity is nothing but the BGW model with infinite number of couplings turned on
in a specific way.
In section \ref{sec:BGW}, we compute the genus expansion of
the BGW free energy using the cut-and-join operator.
We also summarize the known result of the BGW $\tau$-function
and study numerically the large genus behavior of the volume of the moduli space of
super Riemann surfaces.
In section \ref{sec:Bessel}, we consider the Bessel case 
where all couplings are set to zero in the BGW model. We propose a non-perturbative 
completion of the genus expansion of the 
one-point function $\bra Z(\bt)\ket$ in the Bessel case.
In section \ref{sec:low}, we study the low temperature behavior of the
one-point function $\bra Z(\bt)\ket$ in JT supergravity
and find the all-genus result at the lower orders in the low temperature expansion. 
In section \ref{sec:thooft}, we consider
JT supergravity in the large $\nu$ regime
\eqref{eq:thooft-limit} and derive
the constitutive relation for the free energy. 
We also find that there are additional $q$-dependent constants
in the free energy coming from the volume of the unitary group $U(\nu)$.
Finally in section \ref{sec:conclusion} we conclude with some future directions.

\section{\mathversion{bold}JT supergravity and Brezin-Gross-Witten $\tau$-function}\label{sec:JT}
In this section we will show that the connected $n$-point function
$\bra Z(\bt_1)\cdots Z(\bt_n)\ket_c$ in JT supergravity
is written in terms of the Brezin-Gross-Witten (BGW)
$\tau$-function in a specific background.

Let us first recall the result in \cite{Stanford:2019vob}. As explained in \cite{Stanford:2019vob}, the expectation value in the supergravity normalization 
$\bra Z(\bt)\ket_{\text{SJT}}$ and that in the matrix model normalization 
$\bra Z(\bt)\ket_{\text{mat}}$ are related by
\begin{equation}
\begin{aligned}
 \bra Z(\bt)\ket_{\text{SJT}}=\nu+2\bra Z(\bt)\ket_{\text{mat}},
\end{aligned} 
\label{eq:SJT-mat}
\end{equation}
where $\nu$ denotes the RR flux. The first term $\nu$ on the right hand side
of \eqref{eq:SJT-mat} counts the number of supersymmetric ground states, while the factor of $2$ in front of $\bra Z(\bt)\ket_{\text{mat}}$ comes from 
the two-fold degeneracy of the spectrum.
In this paper, we will use the matrix model 
normalization unless otherwise stated.
From the matrix model picture, the non-zero $\nu$
corresponds to the ensemble with $U(N+\nu)\times U(N)$ symmetry.
In the Altland-Zirnbauer classification \cite{altland1997nonstandard} it corresponds to the $(\boldsymbol{\al},\boldsymbol{\bt})=(1+2\nu,2)$ case.
In this ensemble, the supercharge takes the form
\begin{equation}
\begin{aligned}
 Q=\begin{pmatrix} 0& M\\ M^\dagger &0
   \end{pmatrix}
\end{aligned} 
\end{equation}
where $M$ is an $(N+\nu)\times N$ complex matrix.

The correlator $\bra Z(\bt_1)\cdots Z(\bt_n)\ket_c$ in JT supergravity
is constructed by gluing several pieces of building blocks:
the disk and the trumpet partition functions of the super Schwarzian mode, 
and the Weil-Petersson volume of 
the moduli space of super Riemann surfaces. The super Schwarzian partition 
functions on the disk and the 
trumpet are computed in \cite{Stanford:2019vob}
\begin{equation}
\begin{aligned}
Z^{\text{disk}}(\bt,b)&=\hf Z_{\text{SJT}}^{\text{disk}}(\bt)=
e^{S_0}\frac{e^{\frac{\pi^2}{\bt}}}{\rt{2\pi\bt}},\\
Z^{\text{trumpet}}(\bt,b)&=\hf Z_{\text{SJT}}^{\text{trumpet}}(\bt,b)=\hf \frac{e^{-\frac{b^2}{4\bt}}}{\rt{2\pi\bt}},
\end{aligned} 
\label{eq:disk-trumpet}
\end{equation}
where we have used the matrix model normalization. Here and in the rest of this section we set $\nu=0$ for simplicity.

The volume of super moduli space is studied in \cite{norbury2}.
The convention of \cite{norbury2} is different from that in \cite{Stanford:2019vob}
and we
follow the definition in \cite{Stanford:2019vob} in which the volume is 
given by
\begin{equation}
\begin{aligned}
 V_{g,n}(b_1,\cdots,b_n)&=(-1)^n2^{1-g}\int_{\b{\cM}_{g,n}}\Th_{g,n}\exp\left(2\pi^2\ka
+\sum_{i=1}^n\frac{b_i^2}{2}\psi_i\right)\\
&=(-1)^n \int_{\b{\cM}_{g,n}}\Th_{g,n}\exp\left(\pi^2\ka
+\sum_{i=1}^n\frac{b_i^2}{4}\psi_i\right).
\end{aligned} 
\label{eq:Vgn}
\end{equation}
Here $\ka$ is the first Miller-Morita-Mumford class, $\psi_i$ is the first Chern class 
of the line bundle whose fiber is the cotangent space to the marked point $p_i$
on the Riemann surface, $\Th_{g,n}\in H^{4g-4+2n}(\b{\cM}_{g,n})$
is the $\Th$-class introduced by Norbury in \cite{norbury}, and $b_i$ are the lengths
of the geodesic boundaries. 
The geometric meaning of the $\Th$-class is explained in \cite{Witten}.
When going from the first line to the second line of \eqref{eq:Vgn}
we have used the selection rule
\begin{equation}
\begin{aligned}
 \int_{\b{\cM}_{g,n}}\Th_{g,n}\ka^\ell\prod_i\psi_i^{k_i}=0\quad\text{unless}\quad
\ell+\sum_i k_i=g-1.
\end{aligned}
\label{eq:selection} 
\end{equation}
The sign $(-1)^n$ in \eqref{eq:Vgn}
is important as we will see in later sections. Our definition 
\eqref{eq:Vgn} differs from that in \cite{norbury2} by a factor
of $2^{-n}$.

Combining the trumpet partition functions and the Weil-Petersson volume of 
super moduli space,
the $n$-point function for $n\geq3$ is written as
\begin{equation}
\begin{aligned}
 \bra Z(\bt_1)\cdots Z(\bt_n)\ket_c&=\sum_{g=0}^\infty e^{-S_0(2g-2+n)}
\int_0^\infty\prod_{i=1}^n b_idb_i
Z^{\text{trumpet}}(\bt_i,b_i)
V_{g,n}(b_1,\cdots,b_n).
\end{aligned} 
\end{equation}
To simplify the expression, it is convenient to rescale $\bt$ by a factor of $\pi^2$
\begin{equation}
\begin{aligned}
 \bt\to\pi^2\bt.
\end{aligned} 
\label{eq:bt-rescale}
\end{equation}
Then using the selection rule \eqref{eq:selection} we find
\begin{equation}
\begin{aligned}
 \bra Z(\bt_1)\cdots Z(\bt_n)\ket_c=\sum_{g=0}^\infty
\hbar^{2g-2+n}\prod_{i=1}^n\left(\frac{\bt_i}{2\pi}\right)^{\hf}
\int_{\b{\cM}_{g,n}}\Th_{g,n}\frac{e^\ka}{\prod_{i=1}^n (\bt_i\psi_i-1)}.
\end{aligned} 
\label{eq:corr-int}
\end{equation}
Here we have introduced the genus-counting parameter $\hbar$ as
\begin{equation}
\begin{aligned}
 \hbar=\pi e^{-S_0}.
\end{aligned}
\label{eq:def-hbar} 
\end{equation}
For the one- and two-point functions, we need to treat the genus-zero contribution
separately. In the convention \eqref{eq:bt-rescale}
and \eqref{eq:def-hbar}, the genus-zero part of the one-point function, 
i.e. the disk partition function in \eqref{eq:disk-trumpet} is written as
\begin{equation}
\begin{aligned}
 \bra Z(\bt)\ket^{g=0}=\frac{e^{\frac{1}{\bt}}}{\rt{2\pi\bt}\hbar}.
\end{aligned} 
\label{eq:disk}
\end{equation}

The expression \eqref{eq:corr-int} is similar to the bosonic case studied in 
\cite{Okuyama:2019xbv,Okuyama:2020ncd}. In the bosonic JT gravity,
the $n$-point correlator is obtained by
acting the boundary creation operator to the free energy of Kontsevich-Witten (KW)
topological gravity. As emphasized in \cite{Okuyama:2019xbv}, the bosonic 
JT gravity corresponds
to a background of topological gravity where
the infinite number of couplings are turned on with a specific value
\begin{equation}
\begin{aligned}
 t_0=t_1=0,\quad t_n=\frac{(-1)^n}{(n-1)!}~~(n\geq2).
\end{aligned} 
\end{equation}
We will see that a similar relation holds for 
JT supergravity as well.
In the case of JT supergravity, the $n$-point correlator is obtained by
acting the boundary creation operator to the free energy of Brezin-Gross-Witten (BGW)
model \cite{Gross:1980he,Brezin:1980rk,Gross:1991ji,Mironov:1994mv}.
In fact, \eqref{eq:corr-int} is written as
\begin{equation}
\begin{aligned}
 \bra Z(\bt_1)\cdots Z(\bt_n)\ket_c=B(\bt_1)\cdots B(\bt_n)G(1,\{t_k\})\Bigl|_{t_k=0}
\end{aligned} 
\end{equation}
where we introduced the boundary creation operator $B(\bt)$
\begin{equation}
\label{eq:BCO}
\begin{aligned}
 B(\bt)=-\frac{\hbar}{\rt{2\pi}}\sum_{k=0}\bt^{k+\hf}\del_k
\end{aligned} 
\end{equation}
with 
\begin{equation}
\begin{aligned}
 \del_k=\frac{\del}{\del t_k},
\end{aligned} 
\end{equation}
and the generating function of the intersection numbers $G(s,\{t_k\})$
\begin{equation}
\begin{aligned}
 G(s,\{t_k\})&=\sum_{g,n}\frac{\hbar^{2g-2}}{n!}\sum_{\vec{k}\in\mathbb{N}^n}
\int_{\b{\cM}_{g,n}}\Th_{g,n}e^{s\ka}\prod_{i=1}^n\psi_i^{k_i}t_{k_i}.
\end{aligned} 
\end{equation}
As shown in \cite{norbury2}, the $\ka$-class can be replaced by some combination of 
$\psi$-classes and the result of this manipulation is summarized as\footnote{A similar relation for the bosonic case was obtained in \cite{Mulase:2006baa,Dijkgraaf:2018vnm}.}
\begin{equation}
\begin{aligned}
 G(s,\{t_k\})=F(\{t_k+\ga_ks^{k}\}),
\end{aligned} 
\label{eq:G-F}
\end{equation}
where $F(\{t_k\})$ is given by
\begin{equation}
\begin{aligned}
F(\{t_k\})&=\sum_{g,n}\frac{\hbar^{2g-2}}{n!}\sum_{\vec{k}\in\mathbb{N}^n}
\int_{\b{\cM}_{g,n}}\Th_{g,n}\prod_{i=1}^n\psi_i^{k_i}t_{k_i},
\end{aligned} 
\label{eq:F-BGW}
\end{equation}
and $\ga_k$ in \eqref{eq:G-F} is given by
\begin{equation}
\begin{aligned}
 \ga_0=0,\quad
\ga_k=\frac{(-1)^{k-1}}{k!}\quad(k\geq1).
\end{aligned} 
\label{eq:ga-k}
\end{equation}
Finally we arrive at our master equation
\begin{equation}
\begin{aligned}
 \bra Z(\bt_1)\cdots Z(\bt_n)\ket_c=B(\bt_1)\cdots B(\bt_n)F(\{t_k\})\Bigl|_{t_k=\ga_k}.
\end{aligned} 
\end{equation}
In \cite{norbury2} it is proved that $F(\{t_k\})$ in \eqref{eq:F-BGW}
is nothing but the free energy of the BGW 
unitary matrix model \cite{Gross:1980he,Brezin:1980rk,Gross:1991ji,Mironov:1994mv}
\begin{equation}
\begin{aligned}
 e^F=\lim_{N\to\infty}\int_{U(N)}dU e^{\frac{1}{\hbar}\Tr (A^\dag U+A U^\dag)},
\end{aligned} 
\end{equation}
and the coupling $t_k$ is given by the Miwa transform 
\begin{equation}
\begin{aligned}
 t_k=\frac{1}{2k+1}\Tr(A^\dag A)^{-k-\hf}.
\end{aligned} 
\end{equation}
Thus we find that the correlator in JT supergravity is computed from the BGW model
with a specific background $t_k=\ga_k$ with \eqref{eq:ga-k}.
This is analogous to the situation in the bosonic JT gravity as mentioned above.

Note that the disk partition function \eqref{eq:disk} is written as
\begin{equation}
\begin{aligned}
 \bra Z(\bt)\ket^{g=0}=\int_0^\infty dE\rho_0(E)e^{-\bt E},
\end{aligned} 
\end{equation}
where the genus-zero eigenvalue density $\rho_0(E)$ is given by
\begin{equation}
\begin{aligned}
\rho_0(E)=\frac{\cosh2\rt{E}}{\rt{2E}\pi\hbar}. 
\end{aligned} 
\label{eq:rho0}
\end{equation}
From this eigenvalue density $\rho_0(E)$ one can read off the spectral curve 
\cite{Stanford:2019vob}
\begin{equation}
\begin{aligned}
 y=-\frac{\cos(2z)}{\rt{2}z},
\end{aligned} 
\label{eq:curve}
\end{equation}
with $E=-z^2$. From the data of spectral curve, one can compute the
genus expansion of free energy systematically by using the Eynard-Orantin's
topological recursion \cite{Eynard:2007kz}.
Alternatively, 
there is a more direct method to compute the genus expansion, known as the cut-and-join operator which we will review in the next section.

\section{Free energy of Brezin-Gross-Witten model}\label{sec:BGW}
\subsection{Cut-and-join operator}
As shown in \cite{Alexandrov:2016kjl} the free energy of BGW model
is systematically computed by the so-called cut-and-join operator $M$
\begin{equation}
\begin{aligned}
 e^{F(p_i,\hbar)}=e^{\hbar M}\cdot1
\end{aligned} 
\label{eq:CandJ}
\end{equation}
where $M$ is given by \cite{Alexandrov:2016kjl,Do:2016odu}
\begin{equation}
\begin{aligned}
 M=\frac{1-4\nu^2}{8}p_1+\hf\sum_{i,j\in2\mathbb{N}+1}ijp_{i+j+1}\frac{\del^2}{\del p_i\del p_j}
+\sum_{i,j\in2\mathbb{N}+1}(i+j-1)p_ip_j\frac{\del}{\del p_{i+j-1}}.
\end{aligned} 
\label{eq:defM}
\end{equation}
The sum of $i,j$ run over positive odd integers.
$p_{2k+1}$ and the coupling $t_k$ in the previous section are related by
\begin{equation}
\begin{aligned}
 p_{2k+1}=\frac{t_k}{\hbar(2k-1)!!}.
\end{aligned} 
\label{eq:p-t}
\end{equation}

Let us briefly recall the derivation of \eqref{eq:CandJ}.
As we review in section \ref{sec:vir}, the BGW $\tau$-function $\tau(p_i,\hbar)=e^{F(p_i,\hbar)}$
obeys the Virasoro constraint $\mathcal{L}_m\tau(p_i,\hbar)=0$
where $\mathcal{L}_m$ is given by \eqref{eq:Lm}.
Let us consider the following linear combination of the Virasoro constraint
\begin{equation}
\begin{aligned}
 \sum_{m=0}^\infty p_{2m+1}\mathcal{L}_m\tau(p_i,\hbar)=0.
\end{aligned} 
\end{equation}
One can show that this is rewritten as \cite{Alexandrov:2016kjl}
\begin{equation}
\begin{aligned}
 M\tau(p_i,\hbar)=\frac{1}{\hbar}D\tau(p_i,\hbar),
\end{aligned} 
\label{eq:M-D}
\end{equation}
where $M$ is the cut-and-join operator in 
\eqref{eq:defM} and $D$ is the Euler operator
\begin{equation}
\begin{aligned}
 D=\sum_{i\in 2\mathbb{N}+1}ip_i\frac{\del}{\del p_i}.
\end{aligned} 
\end{equation} 
Due to the selection rule \eqref{eq:selection},
$\tau=e^F$ defined by \eqref{eq:F-BGW}
has the property $\tau(p_i,\hbar)=\tau(\hbar^ip_i,1)$.
This means that if we expand $\tau$ as
\begin{equation}
 \tau(p_i,\hbar)=1+\sum_{k=1}^\infty \hbar^k\tau^{(k)}(p_i),
\end{equation}
the coefficients satisfy
\begin{equation}
D\tau^{(k)}(p_i)=k\tau^{(k)}(p_i).
\end{equation}
Therefore using \eqref{eq:M-D} we find
\begin{equation}
\begin{aligned}
 M\tau^{(k)}(p_i)=(k+1)\tau^{(k+1)}(p_i)~~\Rightarrow~~\tau^{(k)}(p_i)=\frac{M^k}{k!}
\cdot1,
\end{aligned} 
\end{equation} 
which implies our desired relation \eqref{eq:CandJ}.

For $\nu=0$, one can compute the first few terms of free energy
by expanding
\eqref{eq:CandJ}
\begin{equation}
\begin{aligned}
 F=\frac{1}{8}p_1\hbar+\frac{1}{16}p_1^2\hbar^2
+\left(\frac{3}{128}p_3+\frac{1}{24}p_1^3\right)\hbar^3
+\left(\frac{9}{128}p_3p_1+\frac{1}{32}p_1^4\right)\hbar^4+\cdots.
\end{aligned}
\label{eq:free-p} 
\end{equation}
Then, from \eqref{eq:p-t} the genus expansion of free energy is obtained as
\begin{equation}
\begin{aligned}
 \sum_{g=0}^\infty \hbar^{2g-2}F_g(\{t_k\})=F\left(p_{2k+1}=\frac{t_k}{\hbar(2k-1)!!},\hbar\right).
\end{aligned} 
\label{eq:rel-cj}
\end{equation} 
Using the dictionary \eqref{eq:rel-cj}
the free energy for $\nu=0$ is obtained from \eqref{eq:free-p} as
\begin{equation}
\begin{aligned}
F_0&=0,\\
F_1&=-\frac{1}{8}\log(1-t_0),\\
F_2&=\frac{3t_1}{128(1-t_0)^3},\\
F_3&=\frac{15t_2}{1024(1-t_0)^5}+\frac{63t_1^2}{1024(1-t_0)^6},\\
F_4&=\frac{2407 t_1^3}{4096(1-t_0)^9}+\frac{8625t_1 t_2}{32768(1-t_0)^8}+
\frac{525t_3}{32768 (1-t_0)^7}.
\end{aligned} 
\label{eq:free-genus}
\end{equation}
Notice that the genus-zero free energy vanishes identically in the BGW model.
This is quite different from the KW case. In particular, we cannot use the KdV equation
to solve the higher genus free energy recursively staring from the 
genus-zero potential $u_0=\del_0^2F_0$, which was 
heavily utilized in the Zograf's approach \cite{Zograf:2008wbe}. 
As we will see in section \ref{sec:thooft}
this problem is avoided
in the large $\nu$ regime \eqref{eq:thooft-limit}.
Note also that the $t_0$-dependence is recovered from the $t_0=0$ result by the 
following rescaling
\begin{equation}
\begin{aligned}
 F_g\to\frac{F_g}{(1-t_0)^{2g-2}},\qquad
t_n\to\frac{t_n}{1-t_0}~~(n\geq1).
\end{aligned} 
\label{eq:t0-scaling}
\end{equation} 

For non-zero $\nu$, the $\nu$-dependence appears through the combination \cite{Alexandrov:2016kjl}
\begin{equation}
\begin{aligned}
 B_k=\prod_{i=1}^k\bigl[(2i-1)^2-4\nu^2\bigr].
\end{aligned} 
\label{eq:Bk}
\end{equation}
For instance, $F_1$ is given by
\begin{equation}
\begin{aligned}
 F_1=-\frac{B_1}{8}\log(1-t_0).
\end{aligned} 
\label{eq:F1-nu}
\end{equation}
The higher genus corrections are easily obtained from the cut-and-join operator
\begin{equation}
\begin{aligned}
 F_2&=\frac{t_1}{384}B_2,\\
F_3&=-\frac{1}{768} t_1^2 B_2+\frac{5 t_1^2+t_2}{15360}B_3,\\
F_4&=\frac{t_1^3}{1152}B_2-\frac{13 t_1^3+3 t_1
   t_2}{9216}B_3+\frac{56 t_1^3+21 t_1
   t_2+t_3}{688128}B_4.
\end{aligned} 
\label{eq:Fg-nu}
\end{equation}
Here we have set $t_0=0$ for simplicity.
The $t_0$-dependence is recovered from the replacement \eqref{eq:t0-scaling}.
\subsection{String equation}
One can also compute the genus expansion using the string equation
for the complex matrix model
\cite{Morris:1990bw,Dalley:1991qg,Dalley:1992br,Klebanov:2003wg}
which reads\footnote{The role of the string equation \eqref{eq:string-eq}
in the context of JT supergravity has been discussed recently
in \cite{Johnson:2019eik,Johnson:2020heh,Johnson:2020exp}.}
\begin{equation}
\begin{aligned}
 \hbar^2\nu^2=\frac{\hbar^2}{4}\left[(\del_0 \cR)^2-2\cR\del_0^2\cR\right]
-2u\cR^2,
\end{aligned} 
\label{eq:string-eq}
\end{equation}
where $\cR$ is a combination of the Gelfand-Dikii differential polynomial
$\cR_k$ 
\begin{equation}
\begin{aligned}
 \cR=\sum_{k=0}^\infty \til{t}_k \cR_k,
\end{aligned} 
\end{equation}
with
\begin{equation}
\begin{aligned}
 \til{t}_k=t_k-\cob_{k,0}.
\end{aligned} 
\end{equation}
Note that $\cR_k$ is normalized as
\begin{equation}
\begin{aligned}
 \cR_0=1,\quad \cR_1=u,\quad \cR_2=\frac{u^2}{2}+\frac{\hbar^2\del_0^2u}{12},\quad
\cR_k=\frac{u^k}{k!}+\cdots,
\end{aligned} 
\end{equation}
where $u$ is defined by
\begin{equation}
\begin{aligned}
 u=\hbar^2\del_0^2F.
\end{aligned} 
\end{equation}
$\cR_k$ can be computed recursively as
\begin{equation}
\begin{aligned}
 (2k+1)\del_0\cR_{k+1}=\frac{\hbar^2}{4}\del_0^3\cR_k+2u\del_0\cR_k+(\del_0u)\cR_k,
\end{aligned} 
\end{equation}
and $u$ obeys the KdV flow equation
\begin{equation}
\begin{aligned}
 \del_ku=\del_0\cR_{k+1}.
\end{aligned} 
\label{eq:kdv}
\end{equation}

As discussed in \cite{Klebanov:2003wg},
the string equation \eqref{eq:string-eq} gives a
non-perturbative definition of the free energy of type 0A minimal superstring theory. 
When the $k$-th coupling is turned on $t_k\ne0$ with higher couplings set to zero
$t_n=0~(n\geq k+1)$, the string equation \eqref{eq:string-eq}
describes the $(2,4k)$ minimal superstring theory.
In this case, the differential equation \eqref{eq:string-eq}
is solved with the boundary condition \cite{Klebanov:2003wg}
\begin{equation}
\begin{aligned}
 u\sim\left\{
\begin{aligned}
 &0,\quad &&(t_0\to\infty),\\
&(1-t_0)^{\frac{1}{k}},\quad &&(t_0\to-\infty).
\end{aligned}
\right.
\end{aligned} 
\end{equation}
In the present case of JT supergravity with the infinite number of couplings 
$t_n=\ga_n$ turned on, it is not clear what is the appropriate boundary condition for
$u$ to solve the string equation \eqref{eq:string-eq}.

However, we are only interested in the formal solution of \eqref{eq:string-eq}
in the small $\hbar$ expansion $u=\sum_{g=0}^\infty \hbar^{2g}u_g$.
For this purpose,
it is not necessary to know the global analytic behavior of $u$ as a function of $t_0$.
It turns out that we can solve the string equation \eqref{eq:string-eq}
order by order in the $\hbar$ expansion just by algebraic manipulations
since $u_g$ is a polynomial in $(1-t_0)^{-1}$.
Let us see this for the first few orders. At the order $\cO(\hbar^0)$,
the string equation \eqref{eq:string-eq} implies $u_0=0$. 
At the next order $\cO(\hbar^2)$, noticing the relation
\begin{equation}
\begin{aligned}
 \cR|_{u_0=0}=t_0-1+\cO(\hbar^2),
\end{aligned} 
\end{equation}
and by equating the order $\cO(\hbar^2)$ terms on the both sides of 
\eqref{eq:string-eq}, we find
\begin{equation}
\begin{aligned}
 \nu^2=\qu-2u_1(t_0-1)^2.
\end{aligned} 
\end{equation}
From the relation $u_1=\del_0^2F_1$,
this reproduces the genus-one free energy $F_1$ in \eqref{eq:F1-nu}.
In this manner we can compute the higher genus free energy $F_g$
by solving the string equation \eqref{eq:string-eq}
order by order in the $\hbar$ expansion. We have checked that
the free energy obtained from \eqref{eq:string-eq} agrees with 
that computed from the cut-and-join operator
in the previous subsection.

\subsection{Free fermion representation of BGW $\tau$-function}\label{sec:vir}
The BGW $\tau$-function satisfies the KdV hierarchy and hence it admits the free fermion 
representation
\begin{equation}
\begin{aligned}
 e^F=\bra t|V\ket.
\end{aligned} 
\end{equation}
Here $\bra t|$ denotes the coherent state of free boson $[\al_n,\al_m]=n\cob_{n+m,0}$
\begin{equation}
\begin{aligned}
\bra t|=\bra 0|\exp\left(\sum_{n\geq0}\frac{t_n}{\hbar(2n+1)!!}\al_{2n+1}\right), 
\end{aligned} 
\end{equation}
and $|V\ket$ is expanded in terms of the free fermion obeying 
$\{\psi_r,\psi_s^*\}=\cob_{r+s,0}$
\begin{equation}
\begin{aligned}
 |V\ket=\exp\left(\sum_{m,n\geq0}A_{m,n}\psi_{-m-\hf}\psi^*_{-n-\hf}\right)|0\ket.
\end{aligned} 
\end{equation}
Note that the boson $\al_n$ and the fermion $\psi_r$ are related by the usual bosonization
\begin{equation}
\begin{aligned}
 \al_n=\sum_{r\in\mathbb{Z}+\hf}:\psi_r\psi^{*}_{n-r}:.
\end{aligned} 
\end{equation}
For the BGW model, the coefficient $A_{m,n}$ is explicitly given by \cite{Dubrovin:2018cho}
\begin{equation}
\begin{aligned}
 A_{m,n}&=(-1)^n\left(\frac{\hbar}{2}\right)^{m+n+1}
\sum_{\substack{r\geq m+1,s\geq0\\ r+s=m+n+1}}\frac{r-s}{r+s}\frac{a_r(\nu)a_s(\nu)}{r!s!},\\
a_r(\nu)&=\Bigl(\hf+\nu\Bigr)_{r}\Bigl(\hf-\nu\Bigr)_{r},
\end{aligned} 
\label{eq:Amn}
\end{equation}
where $(x)_n=x(x+1)\cdots (x+n-1)$ is the Pochhammer symbol.
The generating function of $A_{m,n}$
is given by the modified Bessel functions 
$I_\nu(z),K_\nu(z)$ \cite{Dubrovin:2018cho}
\begin{equation}
\begin{aligned}
 &\sum_{m,n\geq0}A_{m,n}z^{-m-1}w^{-n-1}\Big|_{\hbar=1}\\
=&\frac{2e^{z-w}\rt{4\pi^2zw}}{z^2-w^2}
\Bigl(K_\nu(z)w I_\nu'(w)+I_\nu(w)zK_\nu'(z)\Bigr)-\frac{1}{z-w},
\end{aligned} 
\end{equation}
where we set $\hbar=1$ for simplicity. The $\hbar$-dependence can be easily recovered
by multiplying $A_{m,n}$ by $\hbar^{m+n+1}$.

The important property of the state $|V\ket$ is that
it satisfies the Virasoro constraint
\cite{Fukuma:1990jw,Dijkgraaf:1990rs,Dalley:1992br,Johnson:1993vk}
\begin{equation}
\begin{aligned}
 \mathcal{L}_m|V\ket=0\quad(m\geq0),
\end{aligned}
\label{eq:Vir-con} 
\end{equation}
where the Virasoro generator $\mathcal{L}_m$ is given by
\begin{equation}
\begin{aligned}
 \mathcal{L}_m&=-\frac{\al_{2m+1}}{2\hbar}+L_m,\\
L_m&=\qu\sum_{n\in\mathbb{Z}}
:\al_{2m+2n+1}\al_{-2n-1}:
{}+\frac{1-4\nu^2}{16}\cob_{m,0}.
\end{aligned} 
\end{equation}
In terms of the couplings $\{t_n\}$ the Virasoro generator is written as
\begin{equation}
\begin{aligned}
 \mathcal{L}_m=\frac{\hbar^2}{4}\sum_{i+j=m-1}(2i+1)!!(2j+1)!!\del_i\del_j
+\hf\sum_{i\geq0}\frac{(2i+2m+1)!!}{(2i-1)!!}\til{t}_i\del_{i+m}
+\frac{1-4\nu^2}{16}\cob_{m,0}.
\end{aligned} 
\label{eq:Lm}
\end{equation}
One can check the Virasoro constraint \eqref{eq:Vir-con} order by order in 
the $\hbar$-expansion.
To see this let us expand $|V\ket$ as
\begin{equation}
\begin{aligned}
 |V\ket=\sum_{n=0}^\infty\hbar^{n}|V_n\ket.
\end{aligned} 
\end{equation}
From the explicit form of $A_{m,n}$ in \eqref{eq:Amn},
we find the first few terms of $|V_n\ket$
\begin{equation}
\begin{aligned}
|V_0\ket&=|0\ket,\\
 |V_1\ket&=\hf a_1\psi_{-\hf}\psi_{-\hf}^*|0\ket,\\
|V_2\ket&=\frac{1}{8}a_2\left(\psi_{-\frac{3}{2}}\psi_{-\hf}^*
-\psi_{-\hf}\psi_{-\frac{3}{2}}^*\right)|0\ket,\\
|V_3\ket&=\frac{1}{48}\left[a_3(\psi_{-\frac{5}{2}}\psi_{-\hf}^*
+\psi_{-\hf}\psi_{-\frac{5}{2}}^*)-(a_3+a_2a_1)\psi_{-\frac{3}{2}}
\psi_{-\frac{3}{2}}^*\right]|0\ket.
\end{aligned} 
\end{equation}
Using the above expressions we have checked the Virasoro constraint at order 
$\cO(\hbar^n)~(0\leq n\leq 2)$
\begin{equation}
\begin{aligned}
 L_m|V_n\ket-\hf\al_{2m+1}|V_{n+1}\ket=0.
\end{aligned} 
\end{equation}

As an application of this formalism, let us show that
the genus-zero one-point function $\bra Z(\bt)\ket^{g=0}$ is independent of the
RR flux $\nu$. As shown in \cite{Okuyama:2020ncd}, the one-point function 
$\bra Z(\bt)\ket$ is written as
\begin{equation}
\begin{aligned}
 \bra Z(\bt)\ket=\frac{\bra t|\h{Z}(\bt)|V\ket}{\bra t|V\ket},
\end{aligned} 
\end{equation}
where $\h{Z}(\bt)$ is given by
\begin{equation}
\begin{aligned}
 \h{Z}(\bt)=-\frac{1}{\rt{2\pi}}\sum_{n\in\mathbb{Z}}\bt^{n+\hf}
\frac{\al_{2n+1}}{(2n+1)!!}.
\end{aligned} 
\end{equation}
At genus-zero $|V\ket$ can be replaced by 
$|V_0\ket=|0\ket$
\begin{equation}
\begin{aligned}
 \bra Z(\bt)\ket^{g=0}=\frac{\bra t|\h{Z}(\bt)|0\ket}{\bra t|0\ket},
\end{aligned} 
\end{equation}
which shows that $\bra Z(\bt)\ket^{g=0}$ is 
independent of $\nu$. Thus we can use the same expression \eqref{eq:disk}
for the disk partition function for the non-zero $\nu$ case as well.
Similarly, the genus-zero part of the two-point function is independent of $\nu$
\begin{equation}
\begin{aligned}
 \bra Z(\bt_1)Z(\bt_2)\ket_c^{g=0}=\bra 0|\h{Z}(\bt_1)\h{Z}(\bt_1)|0\ket=
\frac{\rt{\bt_1\bt_2}}{2\pi(\bt_1+\bt_2)}.
\end{aligned} 
\end{equation}

\subsection{Large genus asymptotics}
It is interesting to consider the large genus asymptotics of the Weil-Petersson volume 
$V_{g,1}(b)$ of super Riemann surface, from which we can extract a possible
non-perturbative effects in the $\hbar$-expansion. As usual in string perturbation theory,
$V_{g,1}(b)$ grows as $(2g)!$ and the genus expansion is an asymptotic series.
Note that for the bosonic case $V_{g,1}(b)$ behaves in the large $g$ limit 
as \cite{Zograf:2008wbe,Mirzakhani2,Saad:2019lba}
\begin{equation}
\begin{aligned}
 V_{g,1}(b)\sim \frac{2}{(2\pi)^{\frac{9}{2}}}(4\pi^2)^{2g}
\Ga(2g-3/2)\frac{2}{b}\sinh\frac{b}{2},
\end{aligned} 
\end{equation}
while in the super case $V_{g,1}(b)$ behaves as \cite{Stanford:2019vob}
\begin{equation}
\begin{aligned}
 V_{g,1}(b)\sim -\frac{2^3}{(2\pi)^{\frac{7}{2}}}\left(\frac{\pi}{\rt{2}}\right)^{2g}
\Ga(2g-3/2)\frac{4}{b}\sinh\frac{b}{4} .
\end{aligned} 
\label{eq:asy-super}
\end{equation}
In this section we will check \eqref{eq:asy-super} 
and study its higher order corrections
by means of the numerical analysis using the Richardson transformation.

Once we know the genus-$g$ free energy $F_g$ of the BGW model,
the intersection number is computed as
\begin{equation}
\begin{aligned}
\int_{\b{\cM}_{g,1}} \Th_{g,1} e^\ka\psi_1^d
=\del_dF_g\big|_{t_n=\ga_n},
\end{aligned} 
\label{eq:int-number}
\end{equation}
and $V_{g,1}(b)$ is given by
\begin{equation}
\begin{aligned}
 V_{g,1}(b)&=-
\int_{\b{\cM}_{g,1}}\Th_{g,1}\exp\left(\pi^2\ka+\frac{b^2}{4}\psi_1\right)\\
&=-\sum_{d=0}^{g-1}\pi^{2(g-1-d)}
\left(\frac{b^2}{4}\right)^d \int_{\b{\cM}_{g,1}} \Th_{g,1} e^\ka\psi_1^d .
\end{aligned} 
\label{eq:V-vs-int}
\end{equation}
We have computed $F_g$ up to $g=30$ using the cut-and-join operator.
From these data, we can extract the large genus behavior of the
intersection number \eqref{eq:int-number} by the standard technique of the
Richardson transformation.
In order to find the asymptotic value of the series
\begin{equation}
\begin{aligned}
 S_g=s_0+\frac{s_1}{g}+\frac{s_2}{g^2}+\cdots,\quad \lim_{g\to\infty}S_g=s_0,
\end{aligned} 
\end{equation}
it is convenient to construct its $k$-th Richardson transform
\begin{equation}
\begin{aligned}
 S_g^{(k)}=\sum_{n=0}^k \frac{(-1)^{k+n}(g+n)^nS_{g+n}}{n!(k-n)!}.
\end{aligned} 
\end{equation}
Then one can show that $S_g^{(k)}$ converges to $s_0$ much faster than the original series
$S_g$
\begin{equation}
\begin{aligned}
 S_g^{(k)}-s_0=\cO(g^{-k-1}).
\end{aligned} 
\end{equation}

\begin{figure}[thb]
\centering
\includegraphics[width=8cm]{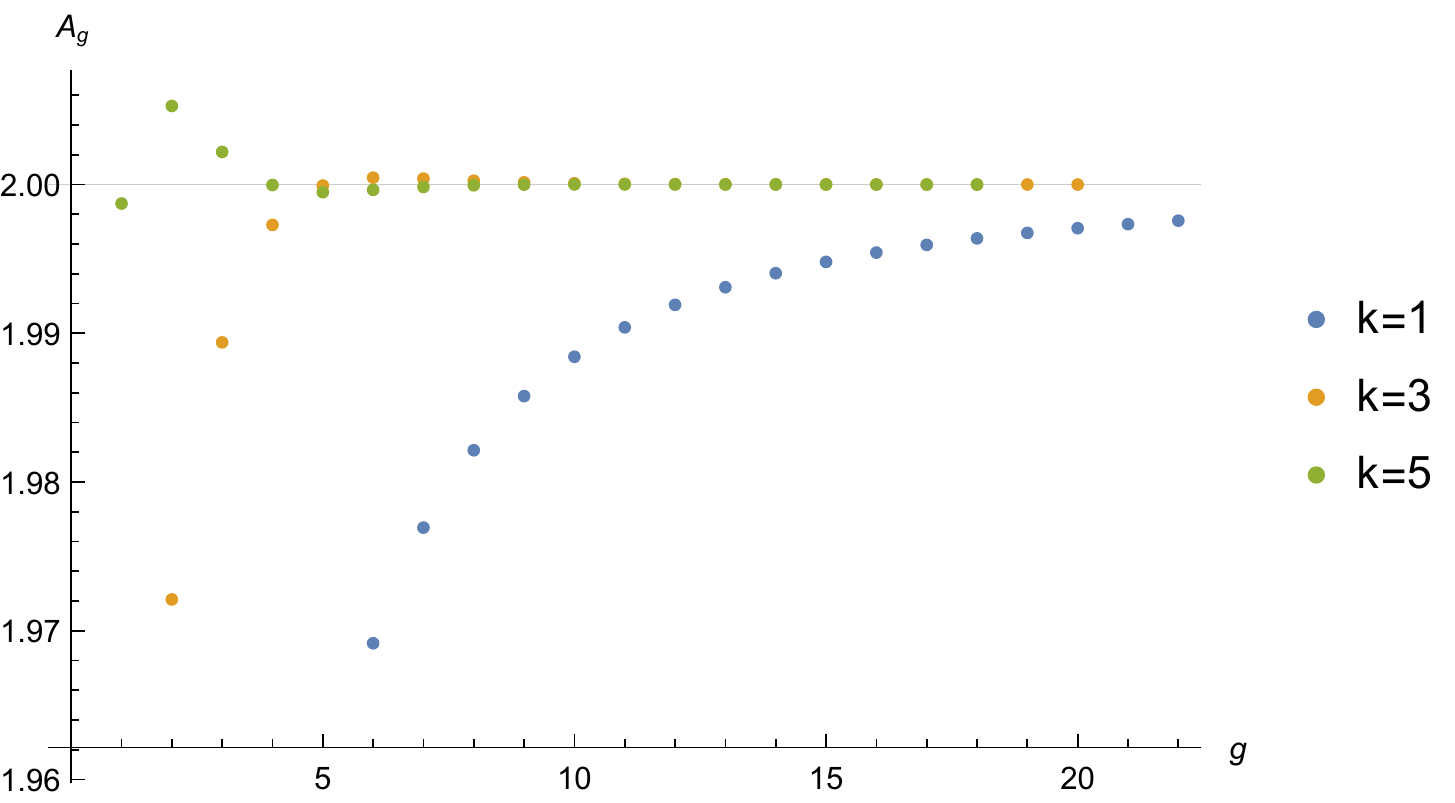}
  \caption{Plot of the $k$-th Richardson transform $A_g^{(k)}$ of $A_g$ in 
\eqref{eq:Ag} for $k=1,3,5$.} 
  \label{fig:richardson}
\end{figure}

Using this acceleration method,
we find the asymptotic behavior of the intersection number
\begin{equation}
\begin{aligned}
\int_{\b{\cM}_{g,1}} \Th_{g,1} e^\ka\psi_1^d \sim 
\rt{\frac{2}{\pi^3}}2^{-g}\frac{\Ga(3/2)}{\Ga(3/2+d)}
\left(\frac{\pi}{4}\right)^{2d}\sum_{n=0}^\infty f_n(d)\Ga(2g-3/2-n)
\end{aligned} 
\label{eq:int-asymp}
\end{equation}
where
\begin{equation}
\begin{aligned}
 f_0=1,\quad f_1=-\frac{1}{8}-\frac{1}{\pi}+\frac{4d-8d^2}{\pi^2}.
\end{aligned} 
\label{eq:f0f1}
\end{equation}
Let us check the behavior \eqref{eq:int-asymp} numerically by 
the Richardson transformation.
For instance, from \eqref{eq:int-asymp} we expect that the combination
\begin{equation}
\begin{aligned}
 A_g=4g^2\frac{\del_0F_g}{\del_0F_{g+1}}\Bigg|_{t_n=\ga_n}
\end{aligned} 
\label{eq:Ag}
\end{equation}
approaches a constant value at large genus
\begin{equation}
\begin{aligned}
 \lim_{g\to \infty}A_g=2.
\end{aligned} 
\end{equation}
As we can see from Fig.~\ref{fig:richardson} the Richardson transformation accelerates 
the convergence of $A_g$ to $2$. We have checked $f_0,f_1$ in \eqref{eq:f0f1} 
in a similar manner by constructing appropriate series 
which converge to $f_0,f_1$ at large genus (see e.g. \cite{Okuyama:2018clk} 
for the details of this procedure).

From \eqref{eq:V-vs-int} we can translate the asymptotic behavior of the intersection number \eqref{eq:int-asymp} to that of the Weil-Petersson volume of super moduli space
\begin{equation}
\begin{aligned}
 V_{g,1}(b)&\sim -
\frac{2^3}{(2\pi)^{\frac{7}{2}}}\left(\frac{\pi}{\rt{2}}\right)^{2g}\Biggl[\Ga(2g-3/2)\frac{4}{b}\sinh\frac{b}{4}\\
&+\Ga(2g-5/2)\left(
-\frac{\sinh
   \left(\frac{b}{4}\right)}{2
   b}-\frac{4 \sinh
   \left(\frac{b}{4}\right)}{b
   \pi }+\frac{8 b \cosh
   \left(\frac{b}{4}\right)-(32+b^2)
   \sinh
   \left(\frac{b}{4}\right)}{2
   b \pi ^2}
\right)+\cdots\Biggr].
\end{aligned} 
\end{equation}
The first line agrees with the result \eqref{eq:asy-super} found in \cite{Stanford:2019vob}, 
while the second line is our new prediction.
Our result implies that there is a non-perturbative correction of the order
\begin{equation}
\begin{aligned}
e^{-\frac{\rt{2}}{\pi}e^{S_0}}=e^{-\frac{\rt{2}}{\hbar}},
\end{aligned} 
\end{equation} 
which is doubly exponential in $S_0$. It would be interesting to 
identify the physical origin of this D-brane like effect
and clarify its meaning in JT supergravity.

\section{Bessel case}\label{sec:Bessel}
If we ignore the $\ka$-class, the BGW model reduces to the so-called
Bessel case corresponding to the spectral curve $y=-\frac{1}{2z}$, or the genus-zero
eigenvalue density $\rho_0(E)=\frac{1}{\rt{2E}\pi\hbar}$.
This density has the behavior $\rho_0(E)\sim E^{-\hf}$ near $E=0$, known as the 
``hard edge'' of the spectrum.
In this section we will first summarize the known results in the Bessel case
and then study the non-perturbative aspects of the
one-point function as well as the spectral form factor
obtained from the two-point function.\footnote{Similar analyses
of the Bessel case in the context of JT supergravity have been done earlier
in \cite{Johnson:2019eik,Johnson:2020heh,Johnson:2020exp},
where nonperturbative definitions of JT gravity and
JT supergravities with/without time-reversal symmetry
are proposed and the connection of the Bessel kernel approach
to the string equation \eqref{eq:string-eq} is also discussed.}

\subsection{One-point function}
The Bessel case corresponds to the trivial background $t_n=0~(n\geq0)$ in the BGW model.
However, it is useful to turn on $t_0\ne0$ and set other couplings $t_n=0~(n\geq1)$.
Then the potential $u=\hbar^2\del_0^2F$ is given by
\begin{equation}
\begin{aligned}
 u(x)=\frac{(1-4\nu^2)\hbar^2}{8x^2},
\end{aligned} 
\end{equation}
where we introduced 
\begin{equation}
\begin{aligned}
 x:=1-t_0.
\end{aligned} 
\end{equation}
The Baker-Akhiezer (BA) function is defined by a solution of the Schr\"{o}dinger equation
\begin{equation}
\begin{aligned}
 \frac{\hbar^2}{2}\del_x^2\psi+u(x)\psi=-E\psi.
\end{aligned} 
\label{eq:schrodinger}
\end{equation}
The solution satisfying 
the boundary condition $\psi(x=0)=0$ is given by the Bessel function
\begin{equation}
\begin{aligned}
\psi= \bra x|E\ket=\frac{\rt{x}}{\hbar}J_\nu \left(\frac{x\rt{2E}}{\hbar}\right).
\end{aligned} 
\label{eq:BA-Bessel}
\end{equation}
This is normalized as
\begin{equation}
\begin{aligned}
 \int_0^\infty dx\bra E|x\ket\bra x|E'\ket=\cob(E-E'),
\end{aligned} 
\end{equation}
which is formally written as
\begin{equation}
\begin{aligned}
 1=\int_0^\infty dx |x\ket\bra x|.
\end{aligned} 
\end{equation}

\begin{figure}[thb]
\centering
\subcaptionbox{$\nu=0$\label{sfig:m0}}{\includegraphics[width=.4\linewidth]{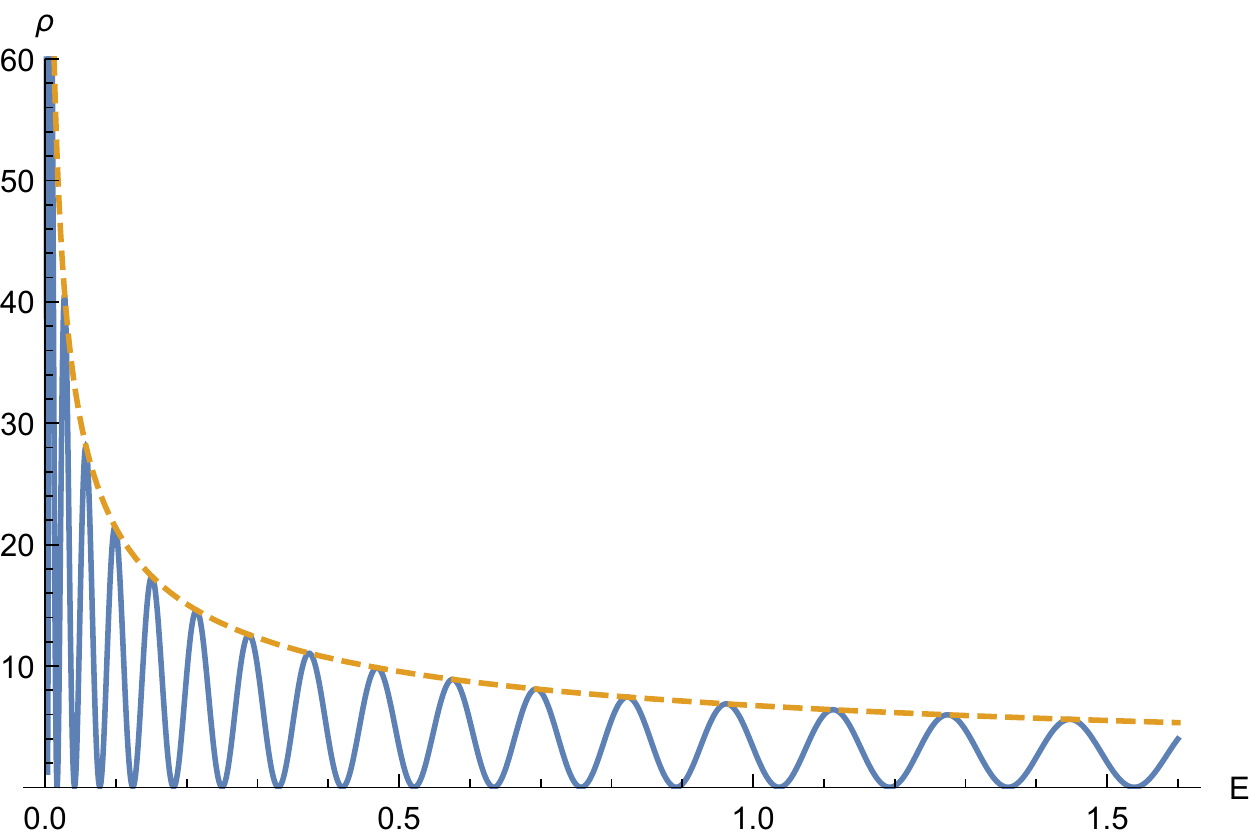}}
\hskip5mm
\subcaptionbox{$\nu=10$\label{sfig:m10}}{\includegraphics[width=.4\linewidth]{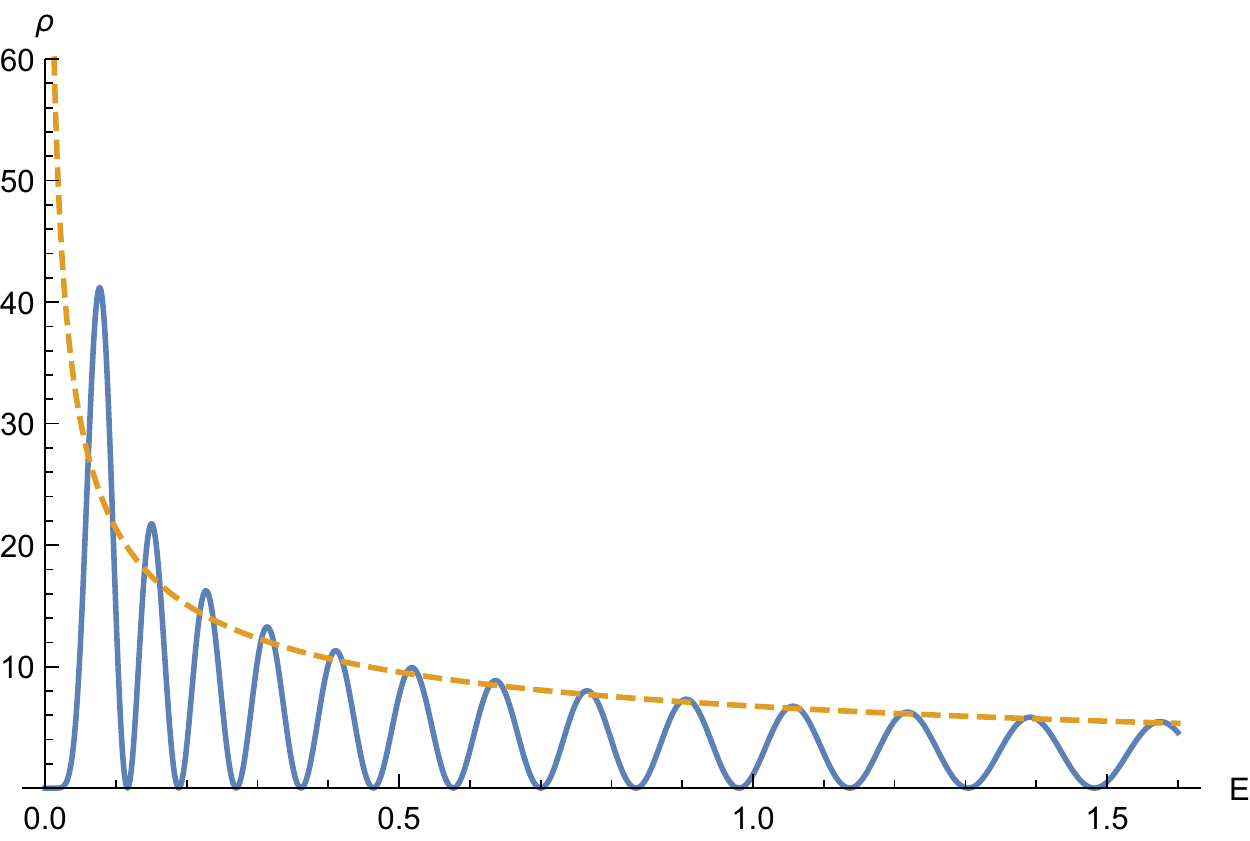}}
  \caption{Plot of the eigenvalue density of 
the Bessel case 
for \subref{sfig:m0} $\nu=0$ and 
\subref{sfig:m10}
$\nu=10$ with $\hbar=1/30$.
The solid curve is the plot of $\rho(E)$ in \eqref{eq:rho-Bessel} while the dashed curve
represents the genus-zero result $\rho_0(E)=\frac{1}{\rt{2E}\pi\hbar}$.} 
  \label{fig:bessel-rho}
\end{figure}

It is also convenient to introduce the projector 
\begin{equation}
\begin{aligned}
 \Pi=\int_0^1 dx |x\ket\bra x|,\qquad \Pi^2=\Pi.
\end{aligned} 
\end{equation}
Then the eigenvalue density is given by \cite{nagao1993nonuniversal}
\begin{equation}
\begin{aligned}
 \rho(E)&=\bra E|\Pi|E\ket=\int_0^1 dx \bra x|E\ket^2
=\frac{J_\nu(\xi)^2-J_{\nu-1}(\xi)J_{1+\nu}(\xi)}{2\hbar^2},
\end{aligned} 
\label{eq:rho-Bessel}
\end{equation}
where
\begin{equation}
\begin{aligned}
 \xi=\frac{\rt{2E}}{\hbar}.
\end{aligned} 
\label{eq:def-xi}
\end{equation}
Note that \eqref{eq:rho-Bessel} is the all-genus result of the eigenvalue density.
See Fig.~\ref{fig:bessel-rho} for the plot of $\rho(E)$
in \eqref{eq:rho-Bessel} for $\nu=0$ and $\nu=10$. One can see that
the genus-zero eigenvalue density $\rho_0(E)\sim E^{-\hf}$ appears as 
the envelope of the oscillating exact eigenvalue density $\rho(E)$.

Using this result, the exact partition function for the Bessel case is computed as
\begin{equation}
\begin{aligned}
 Z(\bt,\nu)&=\int_0^\infty dE\rho(E)e^{-\bt E}\\
&=\frac{(\si/2)^{\nu+1}}{\Ga(2+\nu)}\,
{}_2F_2\Bigl(\nu+\hf,\nu+1;2+\nu,1+2\nu;-2\si\Bigr),
\end{aligned} 
\label{eq:bessel-z}
\end{equation}
where we introduced the parameter $\si$ by
\begin{equation}
\begin{aligned}
 \si=\frac{1}{\hbar^2\bt}.
\end{aligned} 
\label{eq:def-si}
\end{equation}
%For instance
%\begin{equation}
%\begin{aligned}
%2Z(\bt,0)&=e^{-\si}\bigl[I_0(\si)+ I_1(\si)\bigr]\\
% 1+2Z(\bt,1)&=e^{-\si}\bigl[(1+\si)I_0(\si)+\si I_1(\si)\bigr]\\
%2+2Z(\bt,2)&=e^{-\si}\bigl[(2+\si)I_0(\si)+(2+\si)I_1(\si)\bigr]\\
%3+2Z(\bt,3)&=e^{-\si}\bigl[(5+\si)I_0(\si)+(4-4/\si+\si)I_1(\si)\bigr]\\
%4+2Z(\bt,4)&=e^{-\si}\bigl[(8-8/\si+\si)I_0(\si)+(8+16/\si^2-8/\si+\si)I_1(\si)\bigr]
%\end{aligned} 
%\end{equation}
One can check that this obeys a simple relation
\begin{equation}
\begin{aligned}
 2\del_\si  Z(\bt,\nu)=e^{-\si}I_\nu(\si).
\end{aligned} 
\end{equation}
For integer $\nu$, we find
\begin{equation}
\begin{aligned}
 \nu+2Z(\bt,\nu)=e^{-\si}\Bigl[a(\si,\nu)I_\nu(\si)+b(\si,\nu)I_{\nu-1}(\si)\Bigr],
\end{aligned} 
\label{eq:STJ-Bessel}
\end{equation}
where
\begin{equation}
\begin{aligned}
 a(\si,0)&=\si,\quad
&b(\si,0)&=\si,\\
a(\si,1)&=\si,\quad
&b(\si,1)&=\si+1,\\
a(\si,2)&=\si+2,\quad
&b(\si,2)&=\si+4+4\si^{-1},\\
a(\si,3)&=\si+6+6\si^{-1},\quad
&b(\si,3)&=\si+9+24\si^{-1}+24\si^{-2},\\
a(\si,4)&=\si+12+32\si^{-1}+32\si^{-2},\quad
&b(\si,4)&=\si+16+80\si^{-1}+192\si^{-2}+192\si^{-3}.
\end{aligned} 
\end{equation}
Interestingly, the relation \eqref{eq:SJT-mat} between the supergravity
normalization and the matrix model normalization has a natural counterpart in the Bessel case \eqref{eq:STJ-Bessel} as well.

Let us consider the asymptotic genus expansion of the one-point function for $\nu=0$
\begin{equation}
\begin{aligned}
 Z(\bt,0)&=\hf\si e^{-\si}\Bigl[I_0(\si)+I_{1}(\si)\Bigr]\\
&\simeq \frac{1}{\rt{2\pi\bt}\hbar}\left(1-\frac{\hbar^2\bt}{8}-\frac{3\hbar^4\bt^2}{128}-\cdots\right)\\
&=:-\frac{1}{\rt{2\pi}}\sum_{g=0}^\infty c_0(g)(\hbar^2\bt)^{g-\hf},
\end{aligned} 
\label{eq:Z0-Bessel}
\end{equation}
where $c_0(g)$ is given by
\begin{equation}
\begin{aligned}
 c_0(g)=\frac{(2g-1)!!(2g-3)!!}{8^g g!}.
\end{aligned} 
\end{equation}
It is interesting to notice that the coefficients of the genus expansion
in \eqref{eq:Z0-Bessel} are \textit{negative}.
This is consistent with our expression of $\bra Z(\bt)\ket$ in \eqref{eq:corr-int}
for $\ka=0$
\begin{equation}
\begin{aligned}
 Z(\bt,0)=-\rt{\frac{\bt}{2\pi}}\sum_{g\geq0}\hbar^{2g-1}
\int_{\b{\mathcal{M}}_{g,1}}\frac{\Th_{g,1}}{1-\bt\psi_1}.
\end{aligned} 
\end{equation}
Note that this minus sign comes from the definition \eqref{eq:Vgn}
of the Weil-Petersson volume of the super Riemann surface.
Nevertheless, the exact result in the first line of \eqref{eq:Z0-Bessel} 
is a positive function for $\bt>0$.
A non-perturbative resummation of $c_0(g)$ is proposed in
\cite{Bertola:2018emp} in terms of the Tricomi's confluent hypergeometric function, 
which is different from our result \eqref{eq:Z0-Bessel}.
As we have seen above, the integral of the exact eigenvalue density \eqref{eq:rho-Bessel}
uniquely selects the modified Bessel function \eqref{eq:Z0-Bessel}
as the non-perturbative completion.
Note that this type of asymptotic expansion \eqref{eq:Z0-Bessel}
has also appeared in the zero-dimensional Sine-Gordon model \cite{Cherman:2014xia}.

\subsection{Bessel kernel and spectral form factor}
The higher point functions in the Bessel case can in principle
be obtained from a combination of the Darboux-Christoffel kernel
\begin{equation}
\begin{aligned}
 K(E_1,E_2)&=
\bra E_1|\Pi|E_2\ket=
\int_0^1 dx\psi(E_1)\psi(E_2).
\end{aligned} 
\end{equation}
Using the Schr\"{o}dinger equation \eqref{eq:schrodinger} one can show that
\begin{equation}
\begin{aligned}
\lefteqn{(E_1-E_2)K(E_1,E_2)}\\
&=\int_0^1 dx\left[
-\left(\frac{\hbar^2}{2}\del_x^2\psi(E_1)+u\psi(E_1)\right)\psi(E_2)+\psi(E_1)\left(
\frac{\hbar^2}{2}\del_x^2\psi(E_2)+u\psi(E_2)\right)\right]\\
&=\int_0^1 dx\frac{\hbar^2}{2}\del_x \Bigl[\psi(E_1)\del_x\psi(E_2)-\psi(E_2)\del_x\psi(E_1)\Bigr]\\
&=\frac{\hbar^2}{2}\Bigl[\psi(E_1)\del_x\psi(E_2)-\psi(E_2)\del_x\psi(E_1)\Bigr]
\Bigl|_{x=1}.
\end{aligned} 
\end{equation}
Using the explicit form of the BA function \eqref{eq:BA-Bessel} we find 
\begin{equation}
\begin{aligned}
 K(E_1,E_2)=\frac{J_\nu(\xi_1)\xi_2
J_\nu'(\xi_2)-J_\nu(\xi_2)\xi_1
J_\nu'(\xi_1)
}{\hbar^2(\xi^2_1-\xi^2_2)},
\end{aligned} 
\end{equation}
where $\xi$ is defined in \eqref{eq:def-xi}.
This is known as the Bessel kernel. The eigenvalue density is given by
the diagonal part of the Darboux-Christoffel kernel
\begin{equation}
\begin{aligned}
 \rho(E)=K(E,E).
\end{aligned} 
\end{equation}

\begin{figure}[thb]
\centering
\includegraphics[width=8cm]{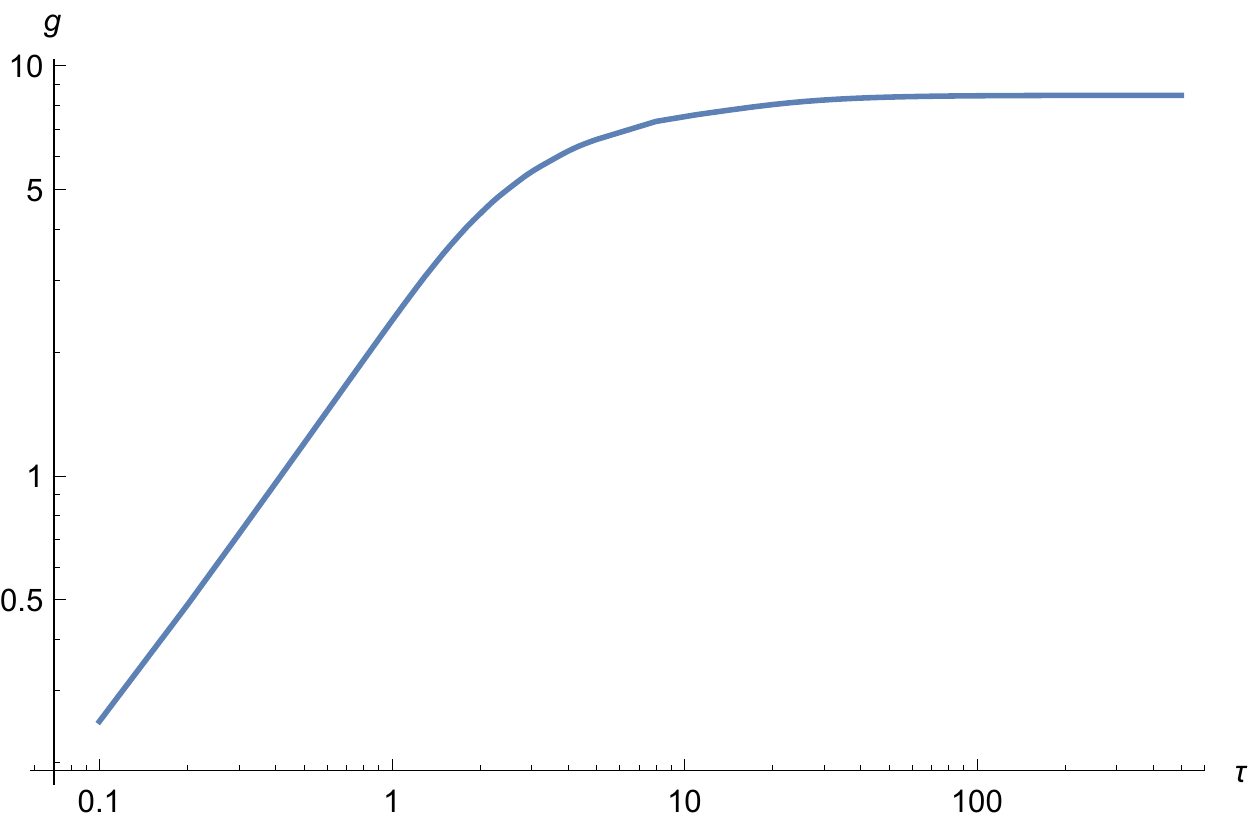}
  \caption{Plot of the spectral form factor for the Bessel case as a function
of $\tau=\hbar t$ for $\hbar=1/30,\bt=1$.} 
  \label{fig:SFF}
\end{figure}

Now let us consider the two-point function in the Bessel case.
The general expression of the two-point function is obtained in \cite{Banks:1989df}
\begin{equation}
\begin{aligned}
 \bra Z(\bt_1)Z(\bt_2)\ket_c=\Tr\bigl[e^{-\bt_1 H}(1-\Pi)e^{-\bt_2 H}\Pi\bigr],
\end{aligned} 
\label{eq:BDSS}
\end{equation}
where $H$ is given by
\begin{equation}
\begin{aligned}
 H=-\frac{\hbar^2}{2}\del_x^2-u(x).
\end{aligned} 
\end{equation}
In terms of the energy eigenstate $H|E\ket=E|E\ket$ \eqref{eq:BDSS} is written as
\begin{equation}
\begin{aligned}
 \bra Z(\bt_1)Z(\bt_2)\ket_c=\int_0^1 dx\int_1^\infty dy \int_0^\infty dE_1\int_0^\infty dE_2 e^{-\bt_1E_1-\bt_2E_2}\bra x|E_1\ket\bra E_1|y\ket\bra y|E_2\ket\bra E_2|x\ket.
\end{aligned} 
\label{eq:ZZ-int}
\end{equation}
Finally, using the Laplace transform of the matrix element
\begin{equation}
\begin{aligned}
 \int_0^\infty dE e^{-\bt E}\bra x|E\ket\bra E|y\ket=\frac{\rt{xy}}{\hbar^2\bt}
e^{-\frac{x^2+y^2}{2\bt \hbar^2}}I_0\left(\frac{xy}{\hbar^2\bt}\right),
\end{aligned} 
\end{equation}
\eqref{eq:ZZ-int} is written as
\begin{equation}
\begin{aligned}
 \bra Z(\bt_1)Z(\bt_2)\ket_c=\int_0^1 dx\int_1^\infty dy
\frac{xy}{\hbar^4\bt_1\bt_2}\exp\left[-\frac{(\bt_1+\bt_2)(x^2+y^2)}{2\hbar^2\bt_1\bt_2}\right]
I_0\left(\frac{xy}{\hbar^2\bt_1}\right)
I_0\left(\frac{xy}{\hbar^2\bt_2}\right).
\end{aligned} 
\label{eq:SFF-Bessel}
\end{equation}
We could not find a closed form expression of this integral.
By numerically evaluating the integral \eqref{eq:SFF-Bessel},
in Fig.~\ref{fig:SFF} we plot the spectral form factor defined by
\begin{equation}
\begin{aligned}
 g(\bt,t,\hbar)=\bra Z(\bt+\ri t)Z(\bt-\ri t)\ket_c.
\end{aligned} 
\end{equation}
One can see from Fig.~\ref{fig:SFF} 
that the spectral form factor exhibits the ramp and the plateau behavior as expected for 
a chaotic system 
\cite{Garcia-Garcia:2016mno,Cotler:2016fpe,Saad:2018bqo}
such as the Sachdev-Ye-Kitaev model \cite{Sachdev,kitaev2015simple} and JT gravity.

\section{Low temperature expansion}\label{sec:low}
Let us go back to the JT supergravity case and consider the one-point function
$\bra Z(\bt)\ket$.
From the result in the Bessel case \eqref{eq:def-si}
it is natural to consider the low temperature regime
\begin{equation}
\begin{aligned}
% \hbar\to0,~~\bt\to\infty\quad\text{with}\quad\si=\frac{1}{\hbar^2\bt}:~\text{fixed}.
\hbar\ll 1,~~\bt\gg 1\quad\text{with}\quad\si=\frac{1}{\hbar^2\bt}:~\text{fixed}.
\end{aligned} 
\label{eq:low}
\end{equation}
A slightly different
low temperature regime (designated as ``the 't Hooft limit'')
was considered in the case of 
bosonic JT gravity in \cite{Okuyama:2019xbv,Okuyama:2020ncd}.
The scale of temperature $T\sim\hbar^{2}$ for $\si\sim\cO(1)$
in \eqref{eq:low}
appears as the characteristic temperature where the contribution from the hard edge
becomes of order one:
\begin{equation}
\begin{aligned}
 \int dE \rho(E)e^{-\bt E}\sim
\int dE \frac{1}{\hbar\rt{E}}e^{-\bt E}\sim\frac{1}{\hbar\bt^{\hf}}\sim\cO(1).
\end{aligned} 
\end{equation}
At present we do not understand the physical origin of this scale
$T\sim\hbar^{2}$  from the bulk JT supergravity picture.
It would be interesting to understand this better.

In this low temperature regime,
the genus-zero part $\bra Z(\bt)\ket^{g=0}$ 
in \eqref{eq:disk} is expanded as
\begin{equation}
\begin{aligned}
 \bra Z(\bt)\ket^{g=0}=\rt{\frac{\si}{2\pi}}e^{\hbar^2\si}=
\frac{1}{\rt{2\pi}}\sum_{\ell=0}^\infty\frac{\hbar^{2\ell}}{\ell!}\si^{\ell+\hf}.
\end{aligned} 
\label{eq:disk-low}
\end{equation}
On the other hand, the contributions from genus $g\geq1$ become
\begin{equation}
\begin{aligned}
 \bra Z(\bt)\ket^{g\geq1}&=-\rt{\frac{\bt}{2\pi}}\sum_{g=1}^\infty \hbar^{2g-1}
\int_{\b{\cM}_{g,1}}\Th_{g,1}\frac{e^\ka}{1-\bt\psi_1}\\
&=-\frac{1}{\rt{2\pi}}\sum_{g=1}^\infty \sum_{\ell=0}^{g-1}\frac{\hbar^{2\ell}}{\ell!}
\si^{\ell+\hf-g}\int_{\b{\cM}_{g,1}}\Th_{g,1}\ka^\ell\psi_1^{g-1-\ell}.
\end{aligned} 
\label{eq:zone-sum}
\end{equation}
Here we have used the selection rule \eqref{eq:selection}.
Thus, for $\nu=0$, we find that the one-point function is expanded as
\begin{equation}
\begin{aligned}
 \bra Z(\bt)\ket=-\frac{1}{\rt{2\pi}}
\sum_{\ell=0}^\infty\sum_{g=0}^{\infty}\hbar^{2\ell}\si^{\ell+\hf-g} 
c_\ell(g).
\end{aligned} 
\label{eq:z-g}
\end{equation}
From \eqref{eq:int-number}, $c_\ell(g)$ is written as
\begin{equation}
\begin{aligned}
 c_\ell(g)=\del_{g-1-\ell}F_g\big|_{t_n=\ga_n}.
\end{aligned} 
\label{eq:cell-Fg}
\end{equation}
We have computed $c_\ell(g)$
using our data of BGW free energy $F_g$ up to $g=30$,
from which we conjecture 
the all genus expression
of $c_\ell(g)$
\begin{equation}
\begin{aligned}
c_0(g)&=\frac{(2g-1)!!(2g-3)!!}{8^g g!},\\
c_1(g)&=\frac{((2g-3)!!)^2}{8^g g!} \frac{ (-1+g) (3-4g+8 g^2)}{3},\\
c_2(g)&=\frac{((2g-5)!!)^2}{8^g g!} \frac{(-2+g) (-1+g)}{180}\\
&\quad\times(-405+2988 g-4176g^2+5632 g^3-4224 g^4+1280g^5),\\
c_3(g)&=\frac{((2g-7)!!)^2}{8^g g!} 
\frac{(-3+g) (-2+g) (-1+g)}{11340}
(70875-108270g+4601928 g^2\\
&\quad
-6412592g^3+7178816 g^4-6335744g^5+3646976 g^6-1132544g^7+143360g^8).
\end{aligned} 
\label{eq:c-ell}
\end{equation}
Although we do not have a proof of \eqref{eq:c-ell},
the agreement of \eqref{eq:c-ell} and 
the derivative of free energy \eqref{eq:cell-Fg} up to $g=30$
gives a strong support for this conjecture \eqref{eq:c-ell}.
Note that the above $c_\ell(g)$ correctly reproduces the
expansion of the disk partition function in \eqref{eq:disk-low} 
\begin{equation}
\begin{aligned}
 -c_\ell(g=0)=\frac{1}{\ell!},
\end{aligned} 
\end{equation}
which is parallel to the case of bosonic JT gravity \cite{Okuyama:2019xbv}.

Let us consider the resummation of $c_\ell(g)$. 
For $\ell=0$, the all-genus resummation of $c_0(g)$ is already obtained
in \eqref{eq:Z0-Bessel} in terms of the modified Bessel function.
We expect that the resummation of $c_\ell(g)$ is expressed as a certain combination of the
modified Bessel functions.
From the asymptotic expansion of the modified Bessel function
\begin{equation}
\begin{aligned}
 I_\nu(z)\approx \frac{e^{z}}{\rt{2\pi z}}\sum_{k=0}^\infty \frac{B_k}{k!(8z)^k},
\end{aligned} 
\label{eq:asymp-Bessel}
\end{equation}
with $B_k$ defined in \eqref{eq:Bk}, 
we find that for $\nu=0$ the one-point function in the low temperature
regime \eqref{eq:low} is expanded as
\begin{equation}
\begin{aligned}
 \bra Z(\bt)\ket=\sum_{\ell=0}^\infty\hbar^{2\ell} e^{-\si}\Bigl[
a_\ell(\si)I_0(\si)+b_\ell(\si)
I_1(\si)\Bigr].
\end{aligned} 
\label{eq:Zab}
\end{equation}
By matching the asymptotic expansion \eqref{eq:asymp-Bessel}
with the all-genus result in \eqref{eq:c-ell}, we find the first few terms of
$a_\ell$ and $b_\ell$
\begin{equation}
\begin{aligned}
a_0&=\frac{\si}{2},\quad &b_0&=\frac{\si}{2},\\
a_1&=\frac{\si}{8}+\si^2-\frac{2
   \si^3}{3},\quad &b_1&=\frac{\si^2}{3}+\frac{2 \si^3}{3},\\
a_2&=\frac{15 \si}{256}-\frac{5
   \si^2}{64}+\frac{137
   \si^3}{120}
&b_2&=
\frac{43
   \si^2}{960}+\frac{\si^3}{40}\\
&\quad
-\frac{21
   \si^4}{5}+\frac{28
   \si^5}{5}-\frac{16 \si^6}{9},\qquad &&\quad
+\frac{31 \si^4}{15}-\frac{212
   \si^5}{45}+\frac{16 \si^6}{9}.
\end{aligned} 
\label{eq:al-bl}
\end{equation}
We emphasize that $a_\ell(\si)$ and $b_\ell(\si)$ contain the all-genus information 
of the intersection numbers $\int_{\b{\cM}_{g,1}}\Th_{g,1}\ka^\ell\psi^{g-1-\ell}$
at the fixed number of $\ka$-insertions.\footnote{We have computed 
$a_\ell(\si),b_\ell(\si)$ up to $\ell=7$. These data 
are available upon request to the authors.}

It is straightforward to generalize this computation to the non-zero $\nu$ case.
For $\nu\ne0$ we find that the one-point function is expanded as
\begin{equation}
\begin{aligned}
 \bra Z(\bt)\ket=\bra Z(\bt)\ket_{\text{Bessel}}+\sum_{\ell=1}^\infty
\hbar^{2\ell} e^{-\si}\Bigl[
a_\ell(\si)I_\nu(\si)+b_\ell(\si)
I_{\nu-1}(\si)\Bigr],
\end{aligned} 
\end{equation}
where the $\ell=0$ term is given by the Bessel case \eqref{eq:bessel-z}.
By matching the asymptotic expansion of $I_\nu(\si),I_{\nu-1}(\si)$ and the genus
expansion of $\bra Z(\bt)\ket$ at $\nu\ne0$, we find
\begin{equation}
\begin{aligned}
 a_1(\si)&=\frac{4 \nu^2-8 \nu+3}{24}\si+\frac{1}{3} (3-2
   \nu)\si^2-\frac{2 \si^3}{3},\\
b_1(\si)&=\frac{1}{3}\si^2+\frac{2\si^3}{3},\\
a_2(\si)&=
\frac{-320
   \nu^5+304 \nu^4+2144 \nu^3-2776
   \nu^2-516 \nu+675}{11520
   }\si\\
&+\frac{560
   \nu^4-2016 \nu^3+1912 \nu^2-72
   \nu-225}{2880}\si^2+\frac{1}{120} \left(92
   \nu^2-248
   \nu+137\right)\si^3\\
&+\frac{1}{45} \left(-40 \nu^2+212
   \nu-189\right)
   \si^4
-\frac{4}{45} (20 \nu-63)
   \si^5-\frac{16 \si^6}{9},\\
b_2(\si)&=
\frac{80
   \nu^4-536 \nu^2+129}{2880
   }\si^2+
\frac{1}{40} \left(28
   \nu^2+1\right)\si^3+\frac{31 \si^4}{15}-\frac{212
   \si^5}{45}
+\frac{16
   \si^6}{9}.
\end{aligned} 
\end{equation}
If we set $\nu=0$ this reduces to \eqref{eq:al-bl} as expected.

In the case of bosonic JT gravity, we observed that the
free energy $\log\bra Z(\bt)\ket$ in the low temperature regime
has an intimate connection to the integral $\int xdy$ on the spectral curve $y=\hf\sin(2z)$ \cite{Okuyama:2019xbv,Okuyama:2020ncd}.
On the other hand, in the present case of JT supergravity we could not find a 
simple relation between the free energy $\log\bra Z(\bt)\ket$ and the spectral curve
$y=-\frac{\cos(2z)}{\rt{2}z}$. 
In the next section, we will see that 
in the large $\nu$ regime \eqref{eq:thooft-limit} 
the relation to the spectral curve becomes
more transparent.

\section{\mathversion{bold}Large $\nu$ regime}\label{sec:thooft}
In this section we consider the large $\nu$ regime \eqref{eq:thooft-limit}. 
This regime was considered before in the BGW model in \cite{Alexandrov:2016kjl} and in the minimal superstring in \cite{Klebanov:2003wg}.
One interesting property of this regime 
is that the free energy of BGW model
admits the genus expansion
\begin{equation}
\begin{aligned}
 F=\sum_{g=0}^\infty\hbar^{2g-2}\cF_g(q,\{t_k\})
\end{aligned} 
\end{equation}
with non-zero genus-zero term $\cF_0\ne0$. This is in contrast to
the case of finite $\nu$ where the genus-zero free energy vanishes $F_0=0$.
We stress that the two statements $F_0=0$ and $\cF_0\ne0$ are not in contradiction.
Let us see this by an example.
In the large $\nu$ limit \eqref{eq:thooft-limit}, the genus-one free
energy $F_1$ in \eqref{eq:F1-nu} becomes
\begin{equation}
\begin{aligned}
 F_1=-\frac{1-4\nu^2}{8}\log(1-t_0)=\frac{q^2}{2\hbar^2}\log(1-t_0)
-\frac{1}{8}\log(1-t_0).
\end{aligned} 
\label{eq:F1-q}
\end{equation}
The first term on the right hand side of \eqref{eq:F1-q} is order $\cO(\hbar^{-2})$
and it can be thought of as a part of the genus-zero free energy $\cF_0$.
Although the genus-zero term vanishes $F_0=0$ in the original
finite $\nu$
case, 
after taking the scaling limit \eqref{eq:thooft-limit},
the order $\nu^{2g}$ term in the genus-$g$ free energy 
$F_g~(g\geq1)$ contributes to the genus-zero 
free energy $\cF_0$ as we have seen in \eqref{eq:F1-q}.

From the result of free energy at non-zero
$\nu$ in \eqref{eq:F1-nu} and \eqref{eq:Fg-nu}, we find that
\begin{equation}
\begin{aligned}
 \cF_0&=\frac{q^2}{2}\log(1-t_0)
+\sum_{\substack{j_a\geq0\\\sum_a j_a=m\geq1\\\sum_a aj_a=n}}\frac{(2n+m-1)!}{(2n+2)!}
\frac{(-q^2)^{n+1}}{2^n(1-t_0)^{2n+m}}\prod_{a=1}^\infty \frac{t_a^{j_a}}{a!^{j_a}j_a!}.
\end{aligned} 
\label{eq:cF0}
\end{equation}
Note that the first term of $\cF_0$ comes from $F_1$ in \eqref{eq:F1-q}
and the rest of \eqref{eq:cF0} comes from $F_g~(g\geq2)$.
This expression
\eqref{eq:cF0} was originally obtained in \cite{Alexandrov:2016kjl}. We will see that
the higher genus corrections $\cF_g~(g\geq1)$ are determined from the 
genus-zero data only, which is known as the constitutive relation \cite{Dijkgraaf:1990nc}. 

Before moving on, let us comment on the
physical meaning of the scaling limit \eqref{eq:thooft-limit}.
As discussed in \cite{Klebanov:2003wg}, in the limit \eqref{eq:thooft-limit}
the factor $\la^\al=\la^{1+2\nu}$ in the measure of the eigenvalue integral
contributes to the effective potential as $e^{-\frac{1}{\hbar}V_{\text{eff}}(\la)}=e^{\frac{2q}{\hbar}\log\la}$, i.e.~the model acquires 
the logarithmic potential like the Penner model.
This modification of the potential
changes the behavior of free energy qualitatively,
e.g.~$F_0=0$ to $\cF_0\ne0$.
We do not understand the bulk gravitational interpretation of this effect.
It would be interesting to clarify the bulk interpretation of this limit
\eqref{eq:thooft-limit}.

\subsection{Genus-zero part}
Let us take a closer look at the genus-zero part \eqref{eq:cF0}.
From \eqref{eq:cF0} the genus-zero potential $u_0=\del_0^2\cF_0$
is given by
\begin{equation}
\label{eq:BGWu0}
\begin{aligned}
 u_0=\sum_{\substack{j_a\geq0\\\sum_a j_a=m\\\sum_a aj_a=n}}\frac{(2n+m+1)!}{(2n+2)!}
\frac{(-q^2)^{n+1}}{2^n(1-t_0)^{2n+m+2}}\prod_{a=1}^\infty \frac{t_a^{j_a}}{a!^{j_a}j_a!}.
\end{aligned} 
\end{equation}
One can check that this satisfies the genus-zero version of the string equation
\eqref{eq:string-eq}
\begin{equation}
\begin{aligned}
 u_0=-\frac{q^2}{2(1-I_0)^2},
\end{aligned} 
\label{eq:BGW-string}
\end{equation}
where we introduced $I_n(u_0)$ by
\begin{equation}
\label{eq:I_n-def}
\begin{aligned}
 I_n(u_0)=\sum_{k=0}^\infty t_{n+k}\frac{u_0^k}{k!}.
\end{aligned} 
\end{equation}
These variables $\{I_n\}$ are introduced by Itzykson and Zuber
\cite{Itzykson:1992ya} in the context of KW model.
Note that in the KW model the genus-zero string equation reads
\begin{equation}
\begin{aligned}
 u_0=I_0(u_0).
\end{aligned}
\label{eq:KW-string} 
\end{equation}

As in the case of KW model, the genus-zero free energy $\cF_0$ in the BGW model
can be written as an integral of $I_0$. To see this let us rewrite
\eqref{eq:BGW-string} as
\begin{equation}
\begin{aligned}
 I_0(u_0)-1+\frac{q}{\rt{-2u_0}}=0.
\end{aligned}
\label{eq:I0-eom} 
\end{equation}
Here we assumed $q>0,u_0<0$ and took the appropriate branch of the square root.
Then we find that $\cF_0$ is written as
\begin{equation}
\begin{aligned}
\cF_0&=\lim_{\ep\to+0}\hf\int_{-\ep}^{u_0}du\Bigl(I_0(u)-1+\frac{q}{\rt{-2u}}\Bigr)^2\\
&\quad +\frac{q^2}{2}\log q-\frac{3}{4}q^2-\frac{q^2}{4}\log\ep.
\end{aligned} 
\label{eq:F0-I0}
\end{equation}
Using the ``equation of motion'' \eqref{eq:I0-eom} one can show that
this expression satisfies the required relation $\del_0^2\cF_0=u_0$
\begin{equation}
\begin{aligned}
 \del_0^2\cF_0=\lim_{\ep\to+0}\int_{-\ep}^{u_0}du=u_0.
\end{aligned} 
\end{equation}
We have also checked that \eqref{eq:F0-I0} reproduces \eqref{eq:cF0} 
in the small $q$ expansion. 
The $q$-dependent constant terms in the second line of \eqref{eq:F0-I0}
are necessary for this agreement, and
we will argue in section \ref{sec:volume} that 
they come 
from the volume of unitary group $U(\nu)$.

Next, let us consider the eigenvalue density at genus-zero from the perspective
of the large $\nu$ regime \eqref{eq:thooft-limit}.\footnote{
A similar derivation of the genus-zero eigenvalue density
in the case of finite $\nu$
(i.e.~without the condition \eqref{eq:thooft-limit})
was considered earlier in \cite{Johnson:2020heh}.
To derive the eigenvalue density correctly, however,
we believe it is essential to consider the large $\nu$ regime
and take the limit of $q\to 0$, as we shall see later.}
At the on-shell value of the coupling $t_n=\ga_n~(n\geq1)$ with $t_0\ne0$, 
the string equation \eqref{eq:BGW-string} is solved as
\begin{equation}
\begin{aligned}
 t_0=J_0(2\rt{u_0})-\frac{q}{\rt{-2u_0}},
\end{aligned} 
\label{eq:onshell-string}
\end{equation}
where we used the relation
\begin{equation}
\begin{aligned}
 \sum_{k=1}^\infty \ga_k\frac{u_0^k}{k!}=
\sum_{k=1}^\infty \frac{(-1)^{k-1}u_0^k}{k!^2}=1-J_0(2\rt{u_0}).
\end{aligned} 
\end{equation}
From this relation \eqref{eq:onshell-string}, $u_0$ is obtained in the small $q$ expansion as
\begin{equation}
\begin{aligned}
 u_0=-\frac{q^2}{2(1-t_0)^2}+\frac{q^4}{2(1-t_0)^5}-\frac{(13+t_0)q^6}{16(1-t_0)^8}+
\cO(q^8).
\end{aligned} 
\label{eq:onshell-u0}
\end{equation} 
Now let us recall that the one-point function is written as the
expectation value of the macroscopic loop operator \cite{Okuyama:2019xbv}
\begin{equation}
\begin{aligned}
 \bra Z(\bt)\ket=\Tr (e^{-\bt H}\Pi)=\int_0^1 dx \bra x|e^{-\bt H}|x\ket.
\end{aligned} 
\label{eq:macro}
\end{equation}
At the genus-zero this reduces to
\begin{equation}
\begin{aligned}
 \bra Z(\bt)\ket^{g=0}&=
\frac{1}{\rt{2\pi\bt}\hbar}\int_0^1 dt_0 e^{\bt u_0}.
\end{aligned} 
\label{eq:z0-t0int}
\end{equation}
As discussed in \cite{Okuyama:2019xbv} it is convenient to change the integration variable
from $t_0$ to $v=-u_0$.
Then using the relation
\begin{equation}
\begin{aligned}
 \int_v^\infty \frac{dE}{\rt{E-v}}e^{-\bt E}=\rt{\frac{\pi}{\bt}}e^{-\bt v},
\end{aligned} 
\end{equation}
\eqref{eq:z0-t0int} is rewritten as the integral of eigenvalue density $\rho(E)$
\begin{equation}
\begin{aligned}
 \bra Z(\bt)\ket^{g=0}
&=\int_{E_0}^\infty dE\rho(E)e^{-\bt E},
\end{aligned} 
\end{equation}
where $\rho(E)$ is given by
\begin{equation}
\begin{aligned}
 \rho(E)&=\frac{1}{\rt{2}\pi\hbar}\int_{E_0}^E dv  \frac{1}{\rt{E-v}}\frac{dt_0}{dv}.
\end{aligned}
\label{eq:rho-int} 
\end{equation}
For $v=-u_0$, $t_0$ in \eqref{eq:onshell-string} becomes
\begin{equation}
\begin{aligned}
 t_0=I_0(2\rt{v})-\frac{q}{\rt{2v}},
\end{aligned} 
\end{equation}
where $I_0(2\rt{v})$ denotes the modified Bessel function of the first kind,
and $E_0$ in \eqref{eq:rho-int} is defined by the zero of $t_0$\footnote{
A similar shift of threshold energy $E_0$ is considered in JT gravity with conical defects \cite{Maxfield:2020ale,WittenJT}.}
\begin{equation}
\begin{aligned}
 I_0(2\rt{E_0})-\frac{q}{\rt{2E_0}}=0.
\end{aligned} 
\end{equation}
Then $\rho(E)$ in \eqref{eq:rho-int} is written more explicitly as 
\begin{equation}
\begin{aligned}
 \rho(E)
&=
\frac{1}{\rt{2}\pi\hbar}\int_{E_0}^E dv \frac{1}{\rt{E-v}}\left(\frac{I_1(2\rt{v})}{\rt{v}}
+\frac{q}{(2v)^{\frac{3}{2}}}\right).
\end{aligned} 
\label{eq:rhoE}
\end{equation}
From \eqref{eq:onshell-u0}, one can easily find the small $q$ 
expansion of $E_0$ as
\begin{equation}
\begin{aligned}
 E_0=-u_0\big|_{t_0=0}=\frac{q^2}{2}-\frac{q^4}{2}+\frac{13q^6}{16}+\cO(q^8).
\end{aligned} 
\end{equation}

Let us consider the $q\to0$ limit of $\rho(E)$ in \eqref{eq:rhoE}.
The first term of \eqref{eq:rhoE} has a naive limit
\begin{equation}
\begin{aligned}
 \frac{1}{\rt{2}\pi\hbar}\int_{0}^E dv \frac{1}{\rt{E-v}}\frac{I_1(2\rt{v})}{\rt{v}}
=\frac{1}{\rt{2}\pi\hbar}\frac{2\sinh^2(\rt{E})}{\rt{E}},
\end{aligned} 
\label{eq:I1-density}
\end{equation}
while the second term of \eqref{eq:rhoE} becomes
\begin{equation}
\begin{aligned}
 \lim_{q\to0}\frac{1}{\rt{2}\pi\hbar}\int_{E_0}^E dv \frac{1}{\rt{E-v}}
\frac{q}{(2v)^{\frac{3}{2}}}&=\lim_{q\to0} \frac{1}{\rt{2}\pi\hbar}\frac{1}{\rt{E}}
\int_{\frac{q^2}{2}} dv \frac{q}{(2v)^{\frac{3}{2}}}
= \frac{1}{\rt{2}\pi\hbar}\frac{1}{\rt{E}}.
\end{aligned} 
\label{eq:1/rtE}
\end{equation}
Putting the two  
contributions \eqref{eq:I1-density} and \eqref{eq:1/rtE} together, we find
\begin{equation}
\begin{aligned}
 \lim_{q\to0}\rho(E)=\frac{\cosh2\rt{E}}{\rt{2E}\pi\hbar},
\end{aligned} 
\end{equation}
which agrees with the genus-zero eigenvalue density $\rho_0(E)$ in 
\eqref{eq:rho0} for $\nu=0$. Thus we showed that the $E^{-\hf}$ hard edge behavior of
the eigenvalue density is correctly reproduced from $\rho(E)$ in \eqref{eq:rhoE}
by carefully taking the limit $q\to0$.
A similar derivation of  $\rho_0(E)$ was considered in \cite{Johnson:2020heh}, 
but the details are different from ours.
In particular, 
neither the large $\nu$ regime nor the limit $q\to0$ were considered
in \cite{Johnson:2020heh} and we believe that 
the appearance of the $E^{-\hf}$ term in \eqref{eq:1/rtE} was not properly explained in 
\cite{Johnson:2020heh}.

\begin{figure}[thb]
\centering
\subcaptionbox{$q=1$\label{sfig:q1}}{\includegraphics[width=.4\linewidth]{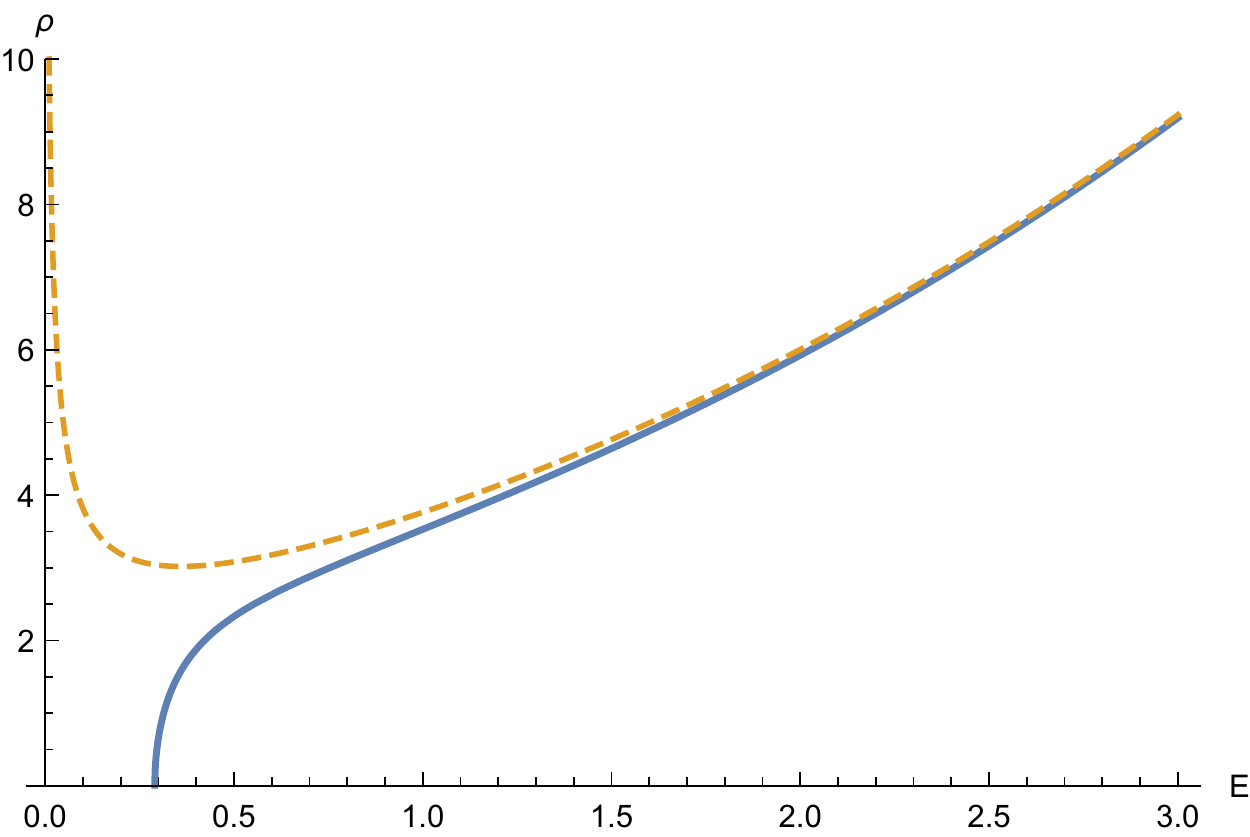}}
\hskip5mm
\subcaptionbox{$q=1/5$\label{sfig:q1/5}}{\includegraphics[width=.4\linewidth]{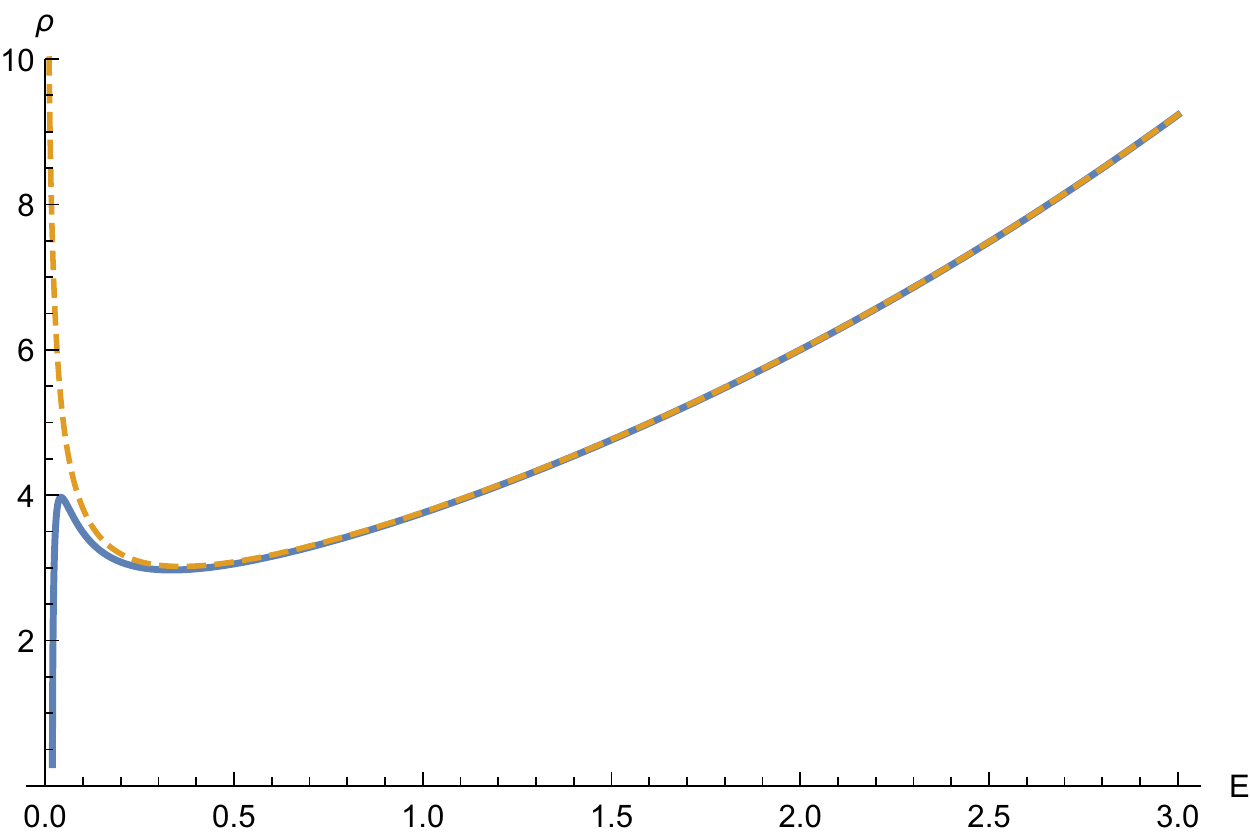}}
  \caption{Plot of the eigenvalue density for \subref{sfig:q1} $q=1$ and \subref{sfig:q1/5}
$q=1/5$.
The solid curve is the plot of $\rt{2}\pi\hbar\rho(E)$ while the dashed curve
represents $\frac{\cosh(2\rt{E})}{\rt{E}}$.} 
  \label{fig:density}
\end{figure}

It is interesting to study the behavior of $\rho(E)$ at non-zero $q$.\footnote{
In a recent paper \cite{Johnson:2020exp}
the eigenvalue density is studied numerically at $\nu=\pm\hf$.
}
In Fig.~\ref{fig:density} we show the plot of \eqref{eq:rhoE} for $q=1$ and $q=1/5$.
At finite $q$, $\rho(E)$ vanishes at $E=E_0$ as 
\begin{equation}
\begin{aligned}
 \lim_{E\to E_0}\rho(E)\sim\rt{E-E_0}.
\end{aligned} 
\label{eq:edge-rho}
\end{equation}
However, as $q\to0$ $\rho(E)$ develops a hard edge $E^{-\hf}$ near $E=0$
as we can see from Fig.~\ref{sfig:q1/5}.

\subsection{Higher genus free energy and constitutive relation}\label{subsec:conrel}

The large $\nu$ BGW free energy $\cF_g$
as well as the KW free energy $F^{\rm KW}_g$
is a formal power series in infinitely many variables $\{t_k\}$.
Itzykson and Zuber found that $F^{\rm KW}_g$
is concisely expressed
when written in the variables $\{I_k\}$ \cite{Itzykson:1992ya}.
It has been known that the free energy can also be expressed concisely
in another set of variables
$\{\partial_0^k u_0\}$ with $u_0=\partial_0^2 F^{\rm KW}_0$
\cite{Eguchi:1994cx,Zhou:2014spa,Zhang:2019hly}.
In other words, higher genus free energies $F^{\rm KW}_g\ (g\ge 1)$
are expressed purely in terms of the genus zero data.
This is known as the constitutive relation \cite{Dijkgraaf:1990nc}.
In what follows we will study the BGW free energy
$\cF_g$ in a similar fashion
and elucidate the constitutive relation for the BGW model.

As mentioned above,
$F^{\rm KW}_g$ are expressed in any of the bases,
$\{t_k\},\ \{I_k\}$ and $\{\partial_0^k u_0\}$.
The explicit transformation formulas among these variables
are known \cite{Zhou:2014spa}.
The advantage of the expressions in the latter two bases is that
one can determine $F^{\rm KW}_g$ iteratively
by merely solving the KdV equation
\begin{align}
\label{eq:KdV}
\partial_1 u
=\partial_0\left[\frac{u^2}{2}+\frac{\hbar^2\partial_0^2 u}{12}\right]
\end{align}
with $u=\hbar^2\partial_0^2 F^{\rm KW}$.
Since $\hbar^2\partial_0^2 F^{\rm BGW}$ also satisfies
the same KdV equation,
it is natural to expect that $\cF_g$ can also be determined
in this way if it is expressed in a suitable basis.
Let us therefore assume that $\cF_g$ is expressed
in the basis $\{\partial_0^k u_0\}$ with $u_0$ given in \eqref{eq:BGWu0}
and try to determine it by solving \eqref{eq:KdV}.
This does not necessarily mean that
we have to work out the full change of variables from
$\{t_k\}$ to $\{\partial_0^k u_0\}$.
Since only derivatives with respect to $t_0,t_1$
appear in \eqref{eq:KdV}, for the purpose of calculating
$\cF_g$ we have only to treat $t_0$ and $t_1$ as variables
and can regard $\{t_k\}\ (k\ge 2)$ as parameters.
From this point of view it is sufficient to consider
the change of variables from $(t_0,t_1)$ to 
\begin{align}
\label{eq:yt-def}
u_0=:y,\quad \partial_0 u_0 =: t^{-1}
\end{align}
with $\{t_k\}\ (k\ge 2)$ held fixed.

For our later purposes,
it is convenient to express the genus zero string equation
in the abstract form
\begin{align}
\label{eq:gstreq}
I_0(u_0)=\varphi(u_0).
\end{align}
For the KW and BGW models the function $\varphi(u_0)$ is given
respectively by
\begin{align}
\label{eq:phiKWBGW}
\varphi_{\rm KW}(u_0)=u_0,\qquad
\varphi_{\rm BGW}(u_0)=1-\frac{q}{\sqrt{-2u_0}}.
\end{align}
By differentiating \eqref{eq:gstreq} in $t_0$ we have
\begin{align}
1+(\partial_0 u_0) I_1(u_0) = (\partial_0 u_0)\varphi'(u_0).
\end{align}
It then follows that
\begin{align}
\label{eq:t-general}
t=\varphi'(u_0)-I_1(u_0).
\end{align}

We next introduce
\begin{align}
\tB_{-1}(v):=-\varphi(v)+\sum_{k=2}^\infty t_k\frac{v^k}{k!}
\end{align}
and define $\tB_n(v)\ (n\ge 0)$ by the relation
\begin{align}
\label{eq:tB-def}
\tB_n=-\partial_v\tB_{n-1},\quad n\ge 0.
\end{align}
$\tB_n$ are related to $I_n$ in \eqref{eq:I_n-def} as
\begin{align}
\label{eq:tB-Irel}
\begin{aligned}
\tB_{-1}(v)
 &=-\varphi(v)-t_0-t_1v+I_0(v),\\
\tB_0(v)
  &=\varphi'(v)+t_1-I_1(v),\\
\tB_n(v)
 &=(-1)^{n+1}\tI_{n+1}(v)\quad(n\ge 1)
\end{aligned}
\end{align}
with
\begin{align}
\tI_n(v) := I_n(v)-\varphi^{(n)}(v)\quad (n\ge 2).
\end{align}
It is clear either from \eqref{eq:I_n-def}
or from \eqref{eq:tB-def}--\eqref{eq:tB-Irel} that
\begin{align}
\label{eq:diffI}
\partial_v\tI_k(v)=\tI_{k+1}(v).
\end{align}
Note that $\tB_n$ is a natural generalization of
$B_n$ used in our previous work \cite{Okuyama:2019xbv,Okuyama:2020ncd}:
$\tB_n$ satisfies the same relation \eqref{eq:tB-def} as $B_n$
and
$\tB_n$ with $\varphi=\varphi_{\rm KW}$
reduces to $B_n$ by setting $t_k=\frac{(-1)^k}{(k-1)!}\ (k\ge 2)$.
In the same sense $\tI_n\ (n\ge 2)$ is a natural generalization
of $I_n$ for the KW model:
$\tI_n$ satisfies \eqref{eq:diffI} and
$\tI_n$ with $\varphi=\varphi_{\rm KW}$
coincides with $I_n$.

The relations \eqref{eq:gstreq} and \eqref{eq:t-general}
are written in terms of $\tB_n$
as
\begin{align}
\label{eq:t01toyt}
t_1=\tB_0(y)-t,\qquad
t_0=-\tB_{-1}(y)-yt_1.
\end{align}
From these relations one can derive
\begin{align}
\label{eq:partial01}
\partial_0=\frac{1}{t}\left(\partial_y-\tB_1(y)\partial_t\right),\qquad
\partial_1=y\partial_0-\partial_t.
\end{align}
This is formally identical to the change of variables
originally introduced by Zograf \cite{Zograf:2008wbe}
and used in our previous work
\cite{Okuyama:2019xbv,Okuyama:2020ncd}.
However, we stress that it is now valid for both KW and BGW models
with full dependence on $t_k\ (k\ge 2)$ being incorporated.
By rewriting $\partial_0$ in \eqref{eq:partial01} as
\begin{align}
\label{eq:partial0bis}
\partial_0=\frac{1}{t}\left(\partial_y-\tI_2(y)\partial_t\right)
\end{align}
and applying it repeatedly on \eqref{eq:yt-def} one obtains
\begin{align}
\label{eq:du0-tI-ex}
\partial_0^2 u_0=\frac{\tI_2}{t^3},\qquad
\partial_0^3 u_0=\frac{\tI_3}{t^4}+\frac{3\tI_2^2}{t^5},\qquad
\partial_0^4 u_0=\frac{\tI_4}{t^5}+\frac{10\tI_2\tI_3}{t^6}
 +\frac{15\tI_2^3}{t^7},
\qquad\cdots.
\end{align}
In general, one can prove that
\begin{align}
\label{eq:du0-tI-rel}
\partial_0^n u_0
 =\sum_{\substack{m_j\ge 0\\ \sum_{j\ge 1}jm_j=n-1}}
 \frac{(\sum_j(j+1)m_j)!}{\prod_j((j+1)!)^{m_j}m_j!}\cdot
 \frac{\prod_j\tI_{j+1}^{m_j}}{t^{\sum_j(j+1)m_j+1}}.
\end{align}
This is a straightforward generalization of the formula 
for the KW model originally obtained in \cite{Zhou:2014spa}.
Again, this formula is now valid for 
both KW and BGW models at arbitrary values of $t_k$.
Note that \eqref{eq:du0-tI-rel} is inverted as \cite{Zhou:2014spa}
\begin{align}
\label{eq:tI-du0-rel}
\tI_n
 =-\sum_{\substack{m_j\ge 0\\ \sum_{j\ge 1}jm_j=n-1}}
 \frac{(\sum_j(j+1)m_j)!}{\prod_j((j+1)!)^{m_j}m_j!}\cdot
 \frac{\prod_j(-\partial_0^{j+1}u_0)^{m_j}}
      {(\partial_0 u_0)^{\sum_j(j+1)m_j+1}}.
\end{align}

With these preparations we can solve the KdV equation
and calculate $\cF_g$ iteratively
in the same way as in the KW case.
As a demonstration let us calculate the genus one free energy
along the lines of \cite{Itzykson:1992ya}.
By plugging the expansion
\begin{align}
u=\sum_{g=0}^\infty\hbar^{2g}u_g
\end{align}
into the KdV equation \eqref{eq:KdV},
one obtains
\begin{align}
\label{eq:KdVg1}
\partial_0\left[(\partial_1-u_0\partial_0)\partial_0\cF_1
 -\frac{1}{12}\partial_0^2 u_0\right]=0
\end{align}
at the order of $\hbar^2$.
From \eqref{eq:du0-tI-ex} and \eqref{eq:partial0bis}
one sees that
\begin{align}
\partial_0^2u_0
 =\frac{\tI_2(y)}{t^3}
 =\partial_t\left(-\frac{\tI_2(y)}{2t^2}\right)
 =\frac{1}{2}\partial_t\partial_0\log t.
\end{align}
Using this and the second relation in \eqref{eq:partial01}
one can rewrite \eqref{eq:KdVg1} as
\begin{align}
\partial_0\partial_t\partial_0
 \left(\cF_1+\frac{1}{24}\log t\right)=0.
\end{align}
This suggests that
\begin{align}
\cF_1=-\frac{1}{24}\log t+\mbox{const.}
\end{align}
Indeed, by means of small $q$ expansion we explicitly verified that
\begin{align}
\label{eq:F1BGW}
\begin{aligned}
\cF_1
 &=-\frac{1}{24}\log t-\frac{1}{12}\log q\\
 &=\frac{1}{24}\log(\partial_0 u_0)-\frac{1}{12}\log q.
\end{aligned}
\end{align}
Similarly, one can compute
the higher genus free energy
$\cF_g\ (g\ge 2)$ up to an integration constant
which cannot be constrained by the KdV equation.
By means of small $q$ expansion we determined the constant term
and verified that
\begin{align}
\label{eq:F2BGW}
\begin{aligned}
\cF_2
 &=\frac{\tI_4}{1152t^3}+\frac{29\tI_2\tI_3}{5760t^4}
   +\frac{7\tI_2^3}{1440t^5}-\frac{1}{240q^2}\\
 &=\frac{\partial_0^4u_0}{1152(\partial_0u_0)^2}
  -\frac{7\partial_0^2u_0\partial_0^3u_0}{1920(\partial_0u_0)^3}
  +\frac{(\partial_0^2u_0)^3}{360(\partial_0u_0)^4}-\frac{1}{240q^2}.
\end{aligned}
\end{align}

The results \eqref{eq:F1BGW}--\eqref{eq:F2BGW} should be
compared with
the well-known results of the KW free energies \cite{Itzykson:1992ya}
\begin{align}
\begin{aligned}
F^{\rm KW}_1
 &=-\frac{1}{24}\log t,\\
F^{\rm KW}_2
 &=\frac{I_4}{1152t^3}+\frac{29I_2I_3}{5760t^4}
   +\frac{7I_2^3}{1440t^5}.
\end{aligned}
\end{align}
One immediately finds that the expressions of
$\cF_g$ and $F^{\rm KW}_g$ are essentially identical
up to the constant term.
This is expected because
both free energies are determined by the same KdV equation
with the common expression of
differential operators \eqref{eq:partial01}
and all $\tI_k$'s appearing in these expressions are originated from
$\tB_1=\tI_2$ in the expression of $\partial_0$.
We conjecture that this correspondence holds for any $g$.

Assuming that this conjecture holds,
let us discuss the constitutive relation for the BGW model.
Since the transformation formula \eqref{eq:du0-tI-rel}
is common to both models,
$\cF_g$ and $F^{\rm KW}_g$ take,
up to the constant term, exactly the same form
when expressed in the $\{\partial_0^k u_0\}$ basis.
This means that the constitutive relations
for the KW and BGW models are identical.
Our discussion is based only on
the KdV equation \eqref{eq:KdV},
the string equation of the form \eqref{eq:gstreq}
and the assumption that the free energy can be expressed
in the basis $\{\partial_0^k u_0\}$.
Thus, we may conclude that the constitutive relation
for the KW model is not specific to the model,
but rather a universal property of
the tau-function for the KdV hierarchy.

Let us present our conjecture more explicitly.
We conjecture that $\cF_g\ (g\ge 2)$ is written in the form
\begin{align}
\label{eq:FgBGWform}
\begin{aligned}
\cF_g
 &=\sum_{\substack{l_j\ge 0\\ \sum_{j=2}^{3g-2}(j-1)l_j=3g-3}}
   b_{l_2\cdots l_{3g-2}}
   \frac{\tI_2^{l_2}\cdots \tI_{3g-2}^{l_{3g-2}}} 
        {t^{2(g-1)+\sum_{j=2}^{3g-2}l_j}}
 +c_gq^{2-2g}\\
 &=\sum_{\substack{l_j\ge 0\\ \sum_{j=2}^{3g-2}(j-1)l_j=3g-3}}
   a_{l_2\cdots l_{3g-2}}
   \frac{(\partial_0^2u_0)^{l_2}\cdots(\partial_0^{3g-2}u_0)^{l_{3g-2}}} 
        {(\partial_0u_0)^{2(1-g)+\sum_{j=2}^{3g-2}jl_j}}
 +c_gq^{2-2g}.
\end{aligned}
\end{align}
Note that with $\varphi=\varphi_{\rm KW}$ and
without the constant term $c_gq^{2-2g}$,
this is exactly the theorem
proved in \cite{Eguchi:1994cx,Zhang:2019hly} for the KW model.
(The first expression for the KW model was originally conjectured
in \cite{Itzykson:1992ya}.)
Once $\cF_g$ is assumed in this form,
one can fully determine the coefficients
$a_{l_2\cdots l_{3g-2}}$ and $b_{l_2\cdots l_{3g-2}}$
by the KdV equation \eqref{eq:KdV}.
Their values are in fact common to both KW and BGW models.
We will also conjecture about the constant term $c_gq^{2-2g}$
in the next subsection.

Let us make a few comments towards a proof of the conjecture.
In the case of the KW model,
the key ingredients of the proof in \cite{Eguchi:1994cx}
are the degree counting and 
the Virasoro condition ${\cal L}_{-1}e^{F^{\rm KW}}=0$.
In the BGW case the degree counting can be done in the same way 
as follows:
We see from \eqref{eq:rel-cj}--\eqref{eq:free-genus}
that we can assign degree $1-k$ to $t_k-\delta_{k,0}$
so that $\cF_g$ is of degree $3g-3$.
This degree assignment works consistently
in the presence of $\nu$ as well ($\nu$ should be of degree $0$).
By further assigning degree $\frac{3}{2}$ to $\hbar$,
we have
\begin{align}
[q]=\frac{3}{2},\quad
[F^{\rm BGW}]=0,\quad
[\cF_g]=3g-3,\quad
[\partial_0^k u_0]=1-k,\quad
[I_k(u_0)-\delta_{k,0}]=1-k,
\end{align}
where $[\cdot]$ denotes the degree. This is the same degree
assignment as in the case of $F^{\rm KW}$ \cite{Eguchi:1994cx}.
On the other hand, the Virasoro ${\cal L}_{-1}$ condition
ensures that $F^{\rm KW}_g\ (g\ge 1)$ does not depend on $u_0$
when written in the $\{\partial_0^k u_0\}$ basis.
However, such a condition seems to be absent in the BGW model
(see \eqref{eq:Vir-con}).
It is therefore not clear to us how to prove the $u_0$-independence of
${\cal F}_g\ (g\ge 1)$.
If one can prove this $u_0$-independence in some way,
the rest of the structure in \eqref{eq:FgBGWform}
can be shown in the same manner as in \cite{Eguchi:1994cx}
based on the degree counting.

To summarize, many of the results about the KW model
that are obtained by solving the KdV equation
are immediately promoted to the universal results
by the mere replacement
\begin{align}
\label{eq:replacement}
\begin{aligned}
I_0&\to u_0,\quad
1-I_1\to t=\varphi'-I_1,\quad
I_k\to\tI_k\ (k\ge 2),\quad
B_k\to\tB_k\ (k\ge 1).
\end{aligned}
\end{align}
We stress that
the universal results are valid
for both KW and BGW models at arbitrary values of $t_k$.
For instance, we can apply this to the results about
the $n$-point correlators $\bra Z(\bt_1)\cdots Z(\bt_n)\ket_c$
and the Baker-Akhiezer function
obtained in our previous work \cite{Okuyama:2019xbv,Okuyama:2020ncd}.

%%%
\subsection{Volume of $U(\nu)$}\label{sec:volume}
We conjecture that the extra $q$-dependent constant comes
from the volume of the unitary group $U(\nu)$ \cite{Ooguri:2002gx}
\begin{equation}
\begin{aligned}
 -\log\Bigl[\text{vol}\,U(\nu)\Bigr]&=\log \frac{G_2(\nu+1)}{(2\pi)^{\hf \nu(\nu+1)}},
\end{aligned} 
\end{equation}
where $G_2(z)$ denotes the Barnes $G$-function.
In the large $\nu$ regime this is expanded as
\begin{equation}
\begin{aligned}
 -\log\Bigl[\text{vol}\,U(\nu)\Bigr]&=\frac{\nu^2}{2}\log\frac{\nu}{2\pi}-\frac{3}{4}\nu^2
-\frac{1}{12}\log\nu+\zeta'(-1)+\sum_{g\geq2}
\frac{B_{2g}}{2g(2g-2)}\nu^{2-2g}\\
&=:\sum_{g=0}^\infty \hbar^{2g-2}\cF_g^{\text{vol}},
\end{aligned} 
\end{equation}
where the genus-$g$ free energy $\cF_g^{\text{vol}}$ coming from the volume 
of $U(\nu)$ is given by
\begin{equation}
\begin{aligned}
 \cF_0^{\text{vol}}&=\frac{q^2}{2}\log\frac{q}{2\pi\hbar}-\frac{3}{4}q^2,\\
\cF_1^{\text{vol}}&=-\frac{1}{12}\log\frac{q}{\hbar}+\zeta'(-1),\\
\cF_g^{\text{vol}}&=\frac{B_{2g}}{2g(2g-2)}q^{2-2g}\quad(g\geq2).
\end{aligned} 
\end{equation}
Here $B_{2g}$ denotes the Bernoulli number.
One can see that the genus zero part $\cF_0^{\text{vol}}$
agrees with the second line of \eqref{eq:F0-I0}
under the identification $\rt{\ep}=2\pi\hbar$.
We observe that $\cF_g^{\text{vol}}~(g\geq2)$ has a negative power of $q$.
On the other hand, the free energy $\cF_g$ of BGW model in the 
large $\nu$ regime 
\eqref{eq:thooft-limit} is a smooth function of $q$ and it contains 
only the positive power of $q$. What is happening is that the negative power of 
$q$ coming from the constitutive part
$\cF_g^{\text{con}}$ is exactly canceled
by $\cF_g^{\text{vol}}$. It turns out that such a term arises
by substituting the leading term in the small $q$ expansion of $u_0$ 
in \eqref{eq:onshell-u0}
\begin{equation}
\begin{aligned}
 \cF_g^{\text{con}}\Big|_{u_0=-\frac{q^2}{2(1-t_0)^2}}=-\cF_g^{\text{vol}}.
\end{aligned} 
\end{equation}
We have checked this relation up to $g=20$.\footnote{The data of 
$\cF_g^{\text{con}}$ up to $g=20$ are available upon request to the authors.}
Then the free energy of BGW model \eqref{eq:FgBGWform} is free from the negative power of $q$
\begin{equation}
\begin{aligned}
 \cF_g=\cF_g^{\text{con}}+\cF_g^{\text{vol}}.
\end{aligned} 
\end{equation}

\subsection{Correlators at large $\nu$}
From the result of free energy $F_g$ at non-zero $\nu$, one can compute the 
connected correlator $\bra Z(\bt_1)\cdots Z(\bt_n)\ket_c$ as a power series in the small
$q$ expansion. It turns out that the result is nicely organized in terms of the variables
\begin{equation}
\begin{aligned}
 \la_i=\hf q^2\bt_i\quad(i=1,\cdots,n).
\end{aligned} 
\end{equation}
For instance, the one-point function is expanded as
\begin{equation}
\begin{aligned}
 \bra Z(\bt)\ket
&=\frac{1}{\rt{4\pi\la}}\sum_{n,m=0}^\infty\nu^{-(2n-1)}q^{2m}\cZ_{n,m}(\la).
\end{aligned} 
\end{equation}
Using the result of free energy $F_g$ obtained from the cut-and-join operator,
we find the closed form expression of the first few terms of this expansion:
\begin{equation}
\begin{aligned}
 \cZ_{0,0}&=e^{-\la}+\rt{\pi\la}\,\text{Erf}(\rt{\la}),\\
\cZ_{0,1}&=\frac{\la+1}{2\la}e^{-\la},\\
\cZ_{0,2}&=\left(\frac{1}{8\la^2}+\frac{1}{8\la}+\frac{1}{16}
+\frac{\la}{4}\right)e^{-\la},\\
\cZ_{1,0}&=\frac{-3\la+2\la^2}{12}e^{-\la},\\
\cZ_{1,1}&=\left(\frac{5}{8}\la-\frac{2}{3}\la^2+\frac{1}{6}\la^3\right)e^{-\la},\\
\cZ_{2,0}&=\left(-\frac{3
   \la^2}{32}+\frac{107
   \la^3}{240}-\frac{29
   \la^4}{120}+\frac{\la^5}{36}\right)e^{-\la},
\end{aligned} 
\label{eq:Zmn-result}
\end{equation}
where $\text{Erf}(z)$ denotes the error function.
In the rest of this section, we will derive these results from the constitutive relations.

Taking the genus-zero term of \eqref{eq:macro}, 
we find
\begin{equation}
\begin{aligned}
 \bra Z(\bt)\ket=\frac{\nu}{\rt{4\pi\la}}\int_0^1 dt_0 e^{\bt u_0}+\cO(\nu^{-1}),
\end{aligned} 
\label{eq:Zone-u0}
\end{equation}
where $u_0$ satisfies the string equation of BGW model \eqref{eq:BGW-string}.
Plugging the small $q$ expansion of $u_0$ in \eqref{eq:onshell-u0}
into \eqref{eq:Zone-u0}
we find
\begin{equation}
\begin{aligned}
 \int_0^1 dt_0 e^{\bt u_0}&=\int_0^1 dt_0 e^{-\frac{\la}{(1-t_0)^2}}\left[1+\frac{\la q^2}{(1-t_0)^5}+\cO(q^4)\right]\\
&=-\rt{\pi\la}+e^{-\la}+\rt{\pi\la}\text{Erf}(\rt{\la})+q^2\frac{1+\la}{2\la}e^{-\la}+\cO(q^4).
\end{aligned} 
\end{equation}
This reproduces the result of $\cZ_{0,m}$ in \eqref{eq:Zmn-result}
up to a constant shift $-\nu/2$ of $\bra Z(\bt)\ket$, which is related to
the matrix model normalization vs.~the supergravity normalization \eqref{eq:SJT-mat}.

The genus-zero part of two-point function is given by \cite{Moore:1991ir,Ginsparg:1993is}
\begin{equation}
\begin{aligned}
 \bra Z(\bt_1)Z(\bt_2)\ket_c^{g=0}&=\frac{\rt{\bt_1\bt_2}}{2\pi(\bt_1+\bt_2)}
e^{(\bt_1+\bt_2)u_0}\Bigl|_{t_0=0}\\
&=\frac{\rt{\la_1\la_2}}{2\pi(\la_1+\la_2)}e^{-(\la_1+\la_2)}\Big[1+q^2(\la_1+\la_2)
+\cO(q^4)\Big].
\end{aligned} 
\label{eq:Ztwo-g0}
\end{equation}
We have checked that this agrees with the direct calculation using $F_g$.
More generally, 
using the genus-zero part of the KdV flow \eqref{eq:kdv}
\begin{equation}
\begin{aligned}
 \del_ku_0=\del_0\frac{u_0^{k+1}}{(k+1)!},
\end{aligned} 
\label{eq:g0-kdv}
\end{equation}
one can show that the genus-zero $n$-point function  is given by
\begin{equation}
\begin{aligned}
 \bra Z(\bt_1)\cdots Z(\bt_n)\ket_c^{g=0}&=(-\hbar\del_0)^{n-2}
\prod_{i=1}^n\rt{\frac{\bt_i}{2\pi}}
\frac{e^{u_0\sum_{i=1}^n\bt_i}}{\sum_{i=1}^n \bt_i}.
\end{aligned} 
\label{eq:Zn-g0}
\end{equation} 
This is the same as the result of the KW model corresponding to a double-scaled hermitian matrix model.  
Note that \eqref{eq:Zn-g0} is a consequence of 
the genus-zero KdV equation \eqref{eq:g0-kdv}, 
which is common for both the KW model and the BGW model. This is the reason why \eqref{eq:Zn-g0} holds for the BGW model as well.

Next let us consider the genus-one part, where the genus-one free energy is given by
\eqref{eq:F1BGW}
\begin{equation}
\begin{aligned}
 \cF_1=\frac{1}{24}\log(\del_0u_0)+\cF_1^{\text{vol}}.
\end{aligned} 
\label{eq:cF1}
\end{equation}
By acting the boundary creation operator we obtain the genus-one part of 
$\bra Z(\bt)\ket$. From \eqref{eq:Zn-g0} we find
%\begin{equation}
%\begin{aligned}
% Bu_0=-\frac{\hbar}{\rt{2\pi}}\sum_{k=0}^\infty\bt^{k+\hf}\del_ku_0=
%-\hbar\rt{\frac{\bt}{2\pi}}\del_0 \frac{e^{\bt u_0}}{\bt}.
%\end{aligned} 
%\end{equation}
%More generally, acting $n$ boundary creation operators on $u_0$ we find
\begin{equation}
\begin{aligned}
 B(\bt_1)\cdots B(\bt_n)u_0=(-\hbar \del_0)^n\prod_{i=1}^n\rt{\frac{\bt_i}{2\pi}}
\frac{e^{u_0\sum_{i=1}^n\bt_i}}{\sum_{i=1}^n\bt_i}.
\end{aligned} 
\label{eq:B-on-u0}
\end{equation}
Then the genus-one part of the correlator 
\begin{equation}
\begin{aligned}
 \bra Z(\bt_1)\cdots Z(\bt_n)\ket_c^{g=1}=B(\bt_1)\cdots B(\bt_n)\cF_1
\end{aligned} 
\end{equation}
is written as a combination of $B(\bt_{i_1})\cdots B(\bt_{i_k})u_0$ in \eqref{eq:B-on-u0}.
From the logarithmic form of $\cF_1$ \eqref{eq:cF1}, the combinatorics is the same 
as the computation of connected correlators. Thus we find
\begin{equation}
\begin{aligned}
 \left\bra\prod_{i=1}^n Z(\bt_i)\right\ket^{g=1}=\frac{1}{24}\log\left[1+
\frac{1}{\del_0u_0}\sum_{n=1}^\infty
\sum_{i_1<\cdots<i_n}x_{i_1}\cdots x_{i_n}
B(\bt_{i_1})\cdots B(\bt_{i_n})\del_0u_0\right]
\Biggl|_{\cO(x_1\cdots x_n)}.
\end{aligned} 
\label{eq:Z-g1}
\end{equation}
For instance, the genus-one part of the one-point function is given by
\begin{equation}
\begin{aligned}
 \bra Z(\bt)\ket^{g=1}
&=\frac{1}{24\del_0u_0}\del_0 (B(\bt)u_0)\\
&=-\frac{1}{\nu\rt{4\pi\la}}\frac{\la}{12}\left(\frac{\del_0^2u_0}{\del_0u_0}
+\bt \del_0u_0\right)e^{\bt u_0}.
\end{aligned} 
\end{equation}
We have checked that this reproduces the $\cZ_{1,m}$ in \eqref{eq:Zmn-result}.
Plugging the on-shell value of $u_0$ in \eqref{eq:onshell-u0}, one can in principle
compute the correlator at genus-one \eqref{eq:Z-g1} up to any order in the
small $q$ expansion.

Finally, let us consider the genus-two part.
The genus-two free energy \eqref{eq:F2BGW} is rewritten as
\begin{equation}
\begin{aligned}
 \cF_2
&=\frac{1}{\del_0u_0}\del_0\left[\frac{1}{1152}\del_0^2\log(\del_0u_0)-\frac{1}{1920}
\Bigl(\del_0\log(\del_0u_0)\Bigr)^2\right]+\cF_2^{\text{vol}}.
\end{aligned} 
\end{equation}
By acting the boundary creation operator $B(\bt)$ 
the genus-two partition function is written as
\begin{equation}
\begin{aligned}
 B(\bt)\cF_2
&=-\cF_2^{\text{con}}B(\bt)\log(\del_0u_0)-\frac{\del_0^2\log(\del_0u_0)}{960\del_0u_0}
\del_0B(\bt)\log(\del_0u_0)\\
&\quad -\frac{\del_0\log(\del_0u_0)}{960\del_0u_0}\del_0^2B(\bt)\log(\del_0u_0)
+\frac{1}{1152\del_0u_0}\del_0^3B(\bt)\log(\del_0u_0).
\end{aligned} 
\end{equation}
Note that $B(\bt)\log(\del_0u_0)$ is proportional to
the genus-one partition function
\begin{equation}
\begin{aligned}
 \bra Z(\bt)\ket^{g=1}=B(\bt)\cF_1=\frac{1}{24}B(\bt)\log(\del_0u_0).
\end{aligned} 
\end{equation}
Thus we find that the genus-two partition function is written in terms of the genus-one
partition function
\begin{equation}
\begin{aligned}
\bra Z(\bt)\ket^{g=2}&=\left(-24\cF_2^{\text{con}}-\frac{\del_0^2\log(\del_0u_0)}{40\del_0u_0}\del_0-
\frac{\del_0\log(\del_0u_0)}{40\del_0u_0}\del_0^2+\frac{1}{48\del_0u_0}\del_0^3\right)\bra Z(\bt)\ket^{g=1}.
\end{aligned} 
\label{eq:Z-g2}
\end{equation}
Again one can in principle compute $\bra Z(\bt)\ket^{g=2}$ up to any order in the 
small $q$ expansion using \eqref{eq:onshell-u0}.

Note that since the genus expansion of the one-point function in JT
gravity has already been calculated in \cite{Okuyama:2019xbv}, one can
derive \eqref{eq:Z-g1} and \eqref{eq:Z-g2} from the corresponding 
bosonic JT results 
simply by the replacement \eqref{eq:replacement} and the change
of variables \eqref{eq:tI-du0-rel},
as stated in Sec.~\ref{subsec:conrel}.\footnote{One also needs
to change the overall sign of the one-point function because
the boundary creation operator \eqref{eq:BCO} in JT supergravity
has an extra overall sign factor as compared to the JT case.}
We verified that \eqref{eq:Z-g1} and \eqref{eq:Z-g2}
are indeed in agreement with the results obtained
in \cite{Okuyama:2019xbv}.

\section{Conclusions and outlook}\label{sec:conclusion}
In this paper we have studied the genus expansion of JT supergravity
using the relation to the BGW $\tau$-function. We found that the matrix model
of JT supergravity is nothing but the BGW model with infinite number of
couplings turned on with specific values $t_k=\ga_k$ in \eqref{eq:ga-k}.
We have computed the genus expansion at finite RR flux $\nu$
using the cut-and-join operator and considered 
the one-point function $\bra Z(\bt)\ket$ in the low temperature
regime \eqref{eq:low}.
We found that  the result is expanded in terms of 
the Bessel functions \eqref{eq:Zab}, which is a natural generalization
of the Bessel case reviewed in section \ref{sec:Bessel}.
Next we have considered the large $\nu$ regime \eqref{eq:thooft-limit}.
It turns out that the genus zero free energy is non-zero in this regime 
and thus at genus-zero we can immediately write down the connected correlator
$\bra Z(\bt_1)\cdots Z(\bt_n)\ket_c^{g=0}$ in terms of the potential $u_0$ (see
\eqref{eq:z0-t0int} and \eqref{eq:Zn-g0}). We have found
that the hard  edge of the eigenvalue
density $\rho_0(E)\sim E^{-\hf}$ is reproduced by carefully taking the limit
$q\to0$ \eqref{eq:1/rtE}.
We have also found that the free energy of 
the BGW model satisfies the constitutive relation, i.e.~the higher genus free energy
is written as a combination of genus-zero quantities $\del_0^nu_0~(n\geq1)$.
This enables us to compute the higher genus corrections to the 
correlators $\bra Z(\bt_1)\cdots Z(\bt_n)\ket_c$
up to any order in principle.
We obtained the $g=1$ correction in the general form \eqref{eq:Z-g1}.
In fact, the constitutive relation of the BGW model is identical
to that of the KW model.
This means that one can get the higher genus results
immediately from the corresponding results in the bosonic JT case
\cite{Okuyama:2019xbv,Okuyama:2020ncd}
by the replacement \eqref{eq:replacement}.
We demonstrated it for the one-point function at
$g=1$ \eqref{eq:Z-g1} and 
$g=2$ \eqref{eq:Z-g2} as an example.

There are several open questions. 
Since the genus expansion of JT supergravity is an asymptotic series
we expect that there appear non-perturbative effects which are analogue
of the D-branes in minimal superstring theory.
It would be interesting to clarify the spacetime picture of these non-perturbative effects.
It would also be interesting to generalize
our computations to other Altland-Zirnbauer ensembles studied in \cite{Stanford:2019vob}
and consider JT (super)gravity on unorientable surfaces along the lines of our work.
Finally, we would like to understand the physical origin of the constitutive relation more
clearly. 
The universality of this relation means that
bosonic JT gravity and supersymmetric JT gravity
are both characterized by
the same form of the free energy of topological gravity.
For this relation to work, it is essential to 
consider the large $\nu$ regime
\eqref{eq:thooft-limit} whose meaning is not well understood in 
the mathematics literature, although this limit
is natural from the viewpoint of physics \cite{Klebanov:2003wg}. 
It would be interesting to clarify this point further.

\acknowledgments
We would like to thank Clifford V.~Johnson for correspondence.
This work was supported in part by JSPS KAKENHI Grant
Nos.~19K03845 and 19K03856,
and JSPS Japan-Russia Research Cooperative Program.

%%%%%%%%%%%%%%%
\bibliography{paper}
\bibliographystyle{utphys}

\end{document}